\documentstyle[fullpage,psfig]{elsart}

\newcommand{\diver}{\mathop{\rm div}\nolimits}

\newcommand{\sign}{\mathop{\rm sign}\nolimits}
\newcommand{\diag}{\mathop{\rm diag}\nolimits}
\newcommand{\minmod}{\mathop{\rm minmod}\nolimits}

\def\vec#1{{\bf #1}}
\newcommand{\etg}{\epsilon_{t}}
\newcommand{\emg}{\epsilon_{m}}
\newcommand{\ewg}{\epsilon_{\omega}}
\def\dfrac#1#2{{\displaystyle#1\over\displaystyle#2}}

\begin{document}
\begin{frontmatter}
\title{TVD Scheme for the Numerical Simulation of the Axisymmetrical
Selfgravitating MHD Flows}
\author[CSU]{Alexander E. Dudorov\thanksref{EM}},
\author[IPM]{Oleg A. Kuznetsov},
\author[CSU]{Andrey G. Zhilkin}

\address[CSU]{Chelyabinsk State University,\\
129, Br. Kashirinykh, Chelyabinsk, 454021, Russia}

\address[IPM]{Keldysh Institute of Applied Mathematics,\\
4, Miusskaya sq., Moscow, 125047, Russia}

\thanks[EM]{E-mails: dudorov@cgu.chel.su;
kuznecov@spp.keldysh.ru; zhag@cgu.chel.su}

\begin{abstract}
The explicit quasi-monotonic conservative TVD scheme and
numerical method for the solution of the gravitational MHD
equations are developed. The 2D numerical code for the
simulation of multidimensional selfgravitating MHD flows on the
Eulerian cylindrical grid is constructed. The results of test
calculations show that the code has a good mathematical and
computational properties and can be applied to the solution of a
wide class of plasma physics and astrophysics problems. We have
simulated, for example, a collapse of magnetized rotating
protostellar clouds.
\end{abstract}

\end{frontmatter}

\section{Introduction}

Magnetic field plays an important role in the various problems
of plasma physics and astrophysics. It is well known, for
example, that the density of the magnetic energy in the
interstellar clouds is compared with the density of the
turbulent energy and can exceed the density of the thermal
energy (see, e.g., \cite{Spitzer1981}). There are a wide class
of problems dealing with the dynamics of the selfgravitating MHD
flows. Note among others the gravitational collapse of the
interstellar clouds, protostellar clusters and galaxies; origin
of the molecular bipolar outflows and jets from young stars and
galaxy cores; dynamics of the accretion flows and circumstellar
disks.

A variety of finite-difference schemes for the solution of the
gasdynamical equations have been developed during last five
decades. Corresponding 1D, 2D, and 3D numerical codes have been
applied to the simulation of various gasdynamical flows. The
direct extension of the these schemes to the MHD, however, is
quite complex problem due to the specific properties of the MHD
equations (anisotropy, vanishing divergence of the magnetic
field, etc.).

Among the non-monotonic generalization of gasdynamical schemes
for numerical simulation of the MHD flows the Lax-Vendroff
method (and its modifications) is used (see, e.g.,
\cite{Dudorov1990}). This method is subjected to the
non-physical oscillations on strong discontinuities
\cite{RichtmyerMorton1967}. Another prevailing method for the
simulation of the multidimensional selfgravitating MHD flows is
the MHD extension of the SPH method \cite{PhilipsMonaghan1985}.

A finite-difference scheme should approximate correctly the
physical oscillations of the flows containing shock waves,
contact discontinuities and rarefaction waves. The nonlinear
monotonic methods of high resolution were suggested for the
solution of these problems. Historically the first monotonic
method was FCT method \cite{BorisBook1973,BorisBook1976}.
It was developed for of linear advection equation and
for gasdynamical equations. At the present time FCT schemes
are used for the simulations of the MHD flows as well
\cite{DeVore1991}.

There are two basic approaches that are applied for the
construction of the monotonic finite-difference schemes for the
numerical solution of the MHD problems. Nonlinear Riemann
problem is solved in the original Godunov's methods at every
mesh interface and every time step by various iteration
procedures (see, e.g., \cite{DaiWoodward1994}). On the basis of
nonlinear Riemann solver, Dai and Woodward
\cite{DaiWoodward1997} extended the PPM method
\cite{ColellaWoodward1984} there. Powell et al.
\cite{Powell1994,Powell1999} constructed an eight-wave
eigensystem for the approximate Riemann solver. Jiang and Wu
\cite{JiangWu1999} applied a high-order WENO interpolation
scheme to MHD equations. Another approach is splitting the
hyperbolicity matrix onto positive and negative parts and
following solution of the linear Riemann problem
\cite{BrioWu1988,ZacharyColella1992,ZacharyMalagoliColella1994,Koldoba1992,RyuJones1995a,RyuJones1995b,Balsara1998a}.

We have suggested recently a simple TVD scheme for the numerical
solution of the MHD equations that has high resolution in the
smooth regions of the solution (see \cite{Dudorov1999}). This
scheme does not demand the solution of full eigenvalues problem
(i.e. a calculation of all eigenvectors and eigenvalues of the
hyperbolicity matrix), since it uses only the spectral radius.
On the basis of developed TVD scheme we elaborated 2D numerical
method for the solution of the selfgravitational MHD equations
in cylindrical coordinates \cite{Dudorov2000}. This method can
be used for the simulation of isothermal and adiabatic plasma
flows. We have applied it to simulation of the collapse of
magnetized rotating protostellar clouds.

In outline this paper proceeds as follows. In Section~2 we
consider the mathematical problem statement for the simulation
of selfgravitating MHD flows in the frame of axisymmetrical
approximation. In Section 3 the TVD scheme for the numerical
solution of the 1D MHD system is constructed. In Section 4 the
developed 1D TVD scheme is extended to the 2D case and numerical
algorithm for simulation of the 2D axisymmetrical
selfgravitating MHD flows is described. Section 5 gives the
description of 2D MHD numerical code and the results of few test
problems. The results of the numerical simulations of the
magnetized protostellar cloud collapse is presented in Section
6. In the conclusion (Section 7) we discuss briefly the main
properties of the presented numerical code and general
perspectives of its development in future.

\section{Problem statement}

\subsection{Basic equations}

The system of the gravitational MHD reads:

\begin{equation}
\frac{\partial \rho}{\partial t}
+
\nabla_{k} (\rho v^{k})
=
0\,,
\label{EQ_2000}
\end{equation}

\begin{equation}
\frac{\partial}{\partial t}
\bigl (
\rho v^{i}
\bigr )
+
\nabla_{k}
\bigl (
\rho v^{i} v^{k}
+
g^{ik} P
-
\sigma^{ik}
\bigr )
=
-\rho \nabla^{i} \Phi\,,
\label{EQ_2001}
\end{equation}

\begin{equation}
\frac{\partial B^{i}}{\partial t}
-
\Bigl [
\nabla ,
\bigl [
\vec{v}, \vec{B}
\bigr ]
\Bigr ]^{i}
=
0\,,
\label{EQ_2002}
\end{equation}

\begin{eqnarray}
\frac{\partial}{\partial t}
\biggl (
\rho \varepsilon
+
\rho \frac{\vec{v}^{2}}{2}
+
\frac{\vec{B}^{2}}{8 \pi}
\biggr )
+
\nabla_{k}
\biggl \{
\rho v^{k}
\biggl (
\varepsilon
+
\frac{P}{\rho}
+
\frac{\vec{v}^{2}}{2}
\biggr )
+
\Bigl [
\vec{B},
[\vec{v}, \vec{B}]
\Bigr ]^{k}
\biggr \} \nonumber \\
\qquad\qquad\qquad\qquad\qquad\qquad
=
-\rho v^{k} \nabla_{k} \Phi\,,
\label{EQ_2003}
\end{eqnarray}

\begin{equation}
\nabla^{2} \Phi = 4 \pi G \rho\,,
\label{EQ_2004}
\end{equation}
where

\begin{equation}
\sigma^{ik}
=
\frac{B^{i}B^{k}}{4 \pi}
-
g^{ik} \frac{\vec{B}^{2}}{8 \pi}
\label{EQ_2005}
\end{equation}
is the Maxwellian tension tensor, $g^{ik}$ is the metric tensor.
The other variables in this system (and further) have its usual
physical sense. The system (\ref{EQ_2000}--\ref{EQ_2005}) is
written in covariant form. Therefore it can be rewritten easy
for the case of arbitrary curvilinear coordinates.

To close the system (\ref{EQ_2000}--\ref{EQ_2005}) we use the
equation of state of a perfect gas $P=(\gamma-1)\rho\varepsilon$,
where $\gamma$ is the adiabatic index. The equation of energy
(\ref{EQ_2003}) in the case of isothermal plasma should be
excluded and the pressure should be determined from the relation
$P=c_{T}^{2}\rho$, $c_{T}^{2}=RT/\mu$, where $c_{T}$ is the
isothermal sound speed, $\mu$ is the molecular weight.

The collapse of a rotating magnetized protostellar cloud can be
studied in the frame of axisymmetric approximation if the
initial uniform magnetic field $\vec{B}$ is parallel to the
vector of angular velocity $\vec{\Omega}$. In this case we can
use the cylindrical coordinates $(r,\varphi,z)$, axis $z$ being
parallel to the vectors $\vec{B}$ and $\vec{\Omega}$. The
variables in (\ref{EQ_2000}--\ref{EQ_2005}) do not depend on the
azimuthal coordinate $\varphi$, so equations
(\ref{EQ_2000}--\ref{EQ_2003}) can be written then in divergent
form as follows:

\begin{equation}
\frac{\partial \vec{u}}{\partial t}
+
\frac{\partial \vec{F}}{\partial r}
+
\frac{\partial \vec{G}}{\partial z}
=
\vec{R}\,,
\label{EQ_2008}
\end{equation}
where the vectors of the conservative variables $\vec{u}$,
fluxes $\vec{F}$, $\vec{G}$ and sources $\vec{R}$ are determined
by expressions:

\begin{eqnarray}
\qquad\qquad
\vec{u}
& =
& \left\{
r\rho\,, \
r\rho v_{r}\,, \ r^2\rho v_{\varphi}\,, \ r\rho v_{z}\,, \
\right. \nonumber \\
&& \qquad \left.
B_{r}\,, \ B_{\varphi}\,, \ rB_{z}\,, \
r\left(\rho \varepsilon
+
\rho \frac{\vec{v}^{2}}{2}
+
\frac{\vec{B}^{2}}{8 \pi}\right)
\right\}^{T}\,,
\label{EQ_2009A}
\end{eqnarray}

\begin{eqnarray}
\qquad {\vec{F}}
& =
& \left\{
r \rho v_{r}\,, \
r\left(\rho v_{r}^{2} + P
+
\frac{\vec{B}^{2}}{8 \pi}
-
\frac{B_{r}^{2}}{4 \pi}\right)\,, \
r^2 \left(\rho v_{r} v_{\varphi}
- \frac{B_{r}B_{\varphi}}{4 \pi}\right)\,, \
\right. \nonumber \\
&& \qquad
r\left(\rho v_{r}v_{z}
- \frac{B_{r}B_{z}}{4 \pi}\right)\,, \
0\,, \
v_{r}B_{\varphi} - v_{\varphi}B_{z}\,, \
r\left(v_{r}B_{z} - v_{z}B_{r}\right)\,, \nonumber \\
&& \qquad \left.
r\rho v_{r}
\left(
\varepsilon
+
\frac{P}{\rho}
+
\frac{\vec{v}^{2}}{2}
\right)
+
r v_{r}\frac{\vec{B}^{2}}{4 \pi}
-
r\frac{B_{r}}{4 \pi}
\left(\vec{v}\cdot\vec{B}
\right)
\right\}^{T}\,,
\label{EQ_2009}
\end{eqnarray}

\begin{eqnarray}
\qquad \vec{G}
& =
& \left\{
r \rho v_{z}\,, \
r \left(
\rho v_{z} v_{r}
-
\dfrac{B_{z} B_{r}}{4 \pi}
\right)\,, \
r^{2}
\left(
\rho v_{\varphi} v_{z}
-
\dfrac{B_{\varphi} B_{z}}{4 \pi}
\right)\,, \
\right. \nonumber \\
&& \qquad
r \left(
\rho v_{z}^{2} + P
+
\dfrac{\vec{B}^{2}}{8 \pi} - \dfrac{B_{z}^{2}}{4 \pi}
\right)\,, \
v_{z} B_{r} - v_{r} B_{z}\,, \
v_{z} B_{\varphi} - v_{\varphi} B_{z}\,, \
0\,, \
\nonumber \\
&& \qquad \left.
r \rho v_{z}
\left(
\varepsilon + \dfrac{P}{\rho} + \dfrac{\vec{v}^{2}}{2}
\right)
+
r v_{z} \dfrac{\vec{B}^{2}}{4\pi}
-
r \dfrac{B_{z}}{4\pi} (\vec{v} \cdot \vec{B})
\right\}^{T}\,,
\label{EQ_2009B}
\end{eqnarray}

\begin{eqnarray}
\qquad \vec{R}
& =
& \left\{
0 \,, \
\rho v_{\varphi}^{2} + P
+
\dfrac{\vec{B}^{2}}{8 \pi}
- \dfrac{B_{\varphi}^{2}}{4 \pi}
-
r \rho \dfrac{\partial \Phi}{\partial r} \,, \
0 \,, \
- r \rho \dfrac{\partial \Phi}{\partial z} \,, \
\right. \nonumber \\
&& \qquad \left.
0 \,, \
0 \,, \
0 \,, \
- r \rho v_{r} \dfrac{\partial \Phi}{\partial r}
- r \rho v_{z} \dfrac{\partial \Phi}{\partial z}
\right \}^{T}\,.
\label{EQ_2009C}
\end{eqnarray}
Here index $T$ denotes the operation of the transposition.

\subsection{Initial and boundary conditions}

Let us consider the problem of initial
and boundary condition on the example of the interstellar
(protostellar) clouds collapse. Under the condition of axial and
equatorial symmetries we can considered the 2D computational
domain in the cylindrical coordinate system:
$D = (0 \le r \le R, 0 \le z \le Z)$ with
the characteristic sizes $R$ and $Z$. We suppose that
contracting cloud is located inside the computational domain and
cloud boundary does not coincide with the grid boundary.

The conditions on the external boundaries of the computational
domain are defined by cloud contraction from the given volume.
In this case the normal components of the velocity $\vec{v}_{n}$
on the external boundary are equal to zero. For density $\rho$,
internal energy $\varepsilon$, magnetic field $\vec{B}$ and
tangential components of the velocity $\vec{v}_{\tau}$ we can
write boundary conditions as $\partial/\partial \vec{n}=0$,
where $\vec{n}$ is the outer normal unit vector. The internal
boundary conditions correspond to the axial and equatorial
symmetries ($v_r=v_\varphi=0$ on the axe and $\partial/\partial
z=0$, $v_z=B_\varphi=B_r=0$ on the equator).

Initial conditions for the collapse of the rotating magnetic
cloud are determined by values of the mass $M_{0}$, temperature
$T_{0}$ and parameters $\etg$, $\emg$, $\ewg$. These parameters
are ratios of the internal, magnetic and kinetic energies of
rotation to the absolute value of gravitational energy,
respectively. The parameters $\etg$, $\emg$, $\ewg$ allow us to
obtain the following similarity relations for the variables
characterizing the initial cloud state:

\begin{equation}
R_{0}
=
8.9111 \cdot 10^{16}
\etg
\biggl (
\frac{M_{0}}{M_{\odot}}
\biggr )
\biggl (
\frac{T_{0}}{10 ~\mathrm{K}}
\biggr )^{-1}
~\mathrm{cm}\,,
\label{EQ_2014}
\end{equation}

\begin{equation}
\rho_{0}
=
6.7475 \cdot 10^{-19}
\etg^{-3}
\biggl (
\frac{M_{0}}{M_{\odot}}
\biggr )^{-2}
\biggl (
\frac{T_{0}}{10 ~\mathrm{K}}
\biggr )^{3}
~\mathrm{g} \cdot \mathrm{cm}^{-3}\,,
\label{EQ_2015}
\end{equation}

\begin{equation}
\Omega_{0}
=
7.5204 \cdot 10^{-13}
\etg^{-3/2}
\ewg^{1/2}
\biggl (
\frac{M_{0}}{M_{\odot}}
\biggr )^{-1}
\biggl (
\frac{T_{0}}{10 ~\mathrm{K}}
\biggr )^{3/2}
~\mathrm{s}^{-1}\,,
\label{EQ_2016}
\end{equation}

\begin{equation}
B_{0}
=
1.2342 \cdot 10^{-4}
\etg^{-2}
\emg^{1/2}
\biggl (
\frac{M_{0}}{M_{\odot}}
\biggr )^{-1}
\biggl (
\frac{T_{0}}{10 ~\mathrm{K}}
\biggr )^{2}
~\mathrm{G}\,.
\label{EQ_2017}
\end{equation}
These similarity relations allow to generalize the results of
simulation with the specified values $\etg$, $\emg$, $\ewg$ on
the contracting clouds with other values of mass and
temperature.

The minimal initial values of the parameters $\etg$, $\emg$,
$\ewg$ depend on the critical mass $M_{*}$. We can obtain the
value of critical mass from virial theorem as follows:

\[
\frac{M_{*}}{M_{0}}
=
\ewg + \frac{3}{2} (\gamma - 1) \etg
+
\sqrt{(\ewg + \frac{3}{2} (\gamma - 1) \etg)^{2} + \emg}\,.
\]

\section{Finite-difference scheme}

\subsection{TVD schemes}

Let us construct the finite-difference scheme for adiabatical
and isothermal MHD flows using the total variation diminishing
(TVD) principle. TVD principle for schemes approximating the
linear hyperbolic system is an extension of monotonic principle
for the schemes approximating the linear advection equation.

The general idea of this principle consists in the numerical
implementation of the total variation diminishing for the
finite-difference Riemann invariants \cite{Harten1984}. An
efficient way of the monotonic schemes construction was proposed
by Vyaznikov et al. \cite{Vyaznikov1989}. It consists in the
following steps:

\begin{enumerate}

\item The selection of a monotonic `base' scheme of the first
order of approximation.

\item The transformation of the `base' scheme to the high
resolution scheme by addition of the antidiffusional terms.

\item The construction of the antidiffusional coefficients for
monotonicity guarantee.

\end{enumerate}

The constructed TVD scheme should have a high resolution in the
smooth regions of the solution but the order of approximation
can decrease in the regions of large gradients.

\subsection{Advection equation}

To investigate the general ideology of the TVD scheme
construction and its basic properties let us consider the linear
advection equation:

\begin{equation}
\frac{\partial \rho}{\partial t}
+
v \frac{\partial \rho}{\partial x}
=
0\,, \qquad\qquad
v = \mathrm{const}\,,
\label{EQ_3001}
\end{equation}
that is defined in the domain $D = \{t > 0$, $-\infty < x < +
\infty \}$ with the initial values $\rho(x,0)=\rho_{0}(x)$. We
introduce an uniform spatial grid $\{ x_{i} = ih$, $i = 0, \pm
1, \pm 2, \ldots \}$ with the fixed stepsize $h$ and consider
the following class of `base' schemes for the equation
(\ref{EQ_3001}):

\begin{eqnarray}
\frac{\rho_{i}^{n+1}-\rho_{i}^{n}}{\tau}
& +
& v \frac{\rho_{i+1}^{n}-\rho_{i-1}^{n}}{2h}
-
\frac{w}{2h}
\bigl (
\rho_{i+1}^{n}
-
2\rho_{i}^{n}
+
\rho_{i-1}^{n}
\bigr ) \nonumber \\
& -
& \frac{v-w}{4}
\left (
\alpha_{i+1/2}^{-}
\frac{\rho_{i+1}^{n}-\rho_{i}^{n}}{h}
-
\alpha_{i-1/2}^{-}
\frac{\rho_{i}^{n}-\rho_{i-1}^{n}}{h}
\right ) \nonumber \\
& +
& \frac{v+w}{4}
\left (
\alpha_{i+1/2}^{+}
\frac{\rho_{i+1}^{n}-\rho_{i}^{n}}{h}
-
\alpha_{i-1/2}^{+}
\frac{\rho_{i}^{n}-\rho_{i-1}^{n}}{h}
\right )
=
0\,.
\label{EQ_3002}
\end{eqnarray}
The parameter $w$ is a coefficient of numerical viscosity, while
the coefficients $\alpha_{i+1/2} \ge 0$ determine the
antidiffusional terms.

If all $\alpha_{i+1/2} = 0$ the scheme (\ref{EQ_3002}) gives the
`base' scheme of the first order of approximation in space and
time. The condition of monotonicity of this scheme is $w \ge
|v|$. In the case $w = |v|$ the `base' scheme transforms into the
classical donor cell scheme \cite{CourantIsaacsonRees1952}.

To obtain the conditions of monotonicity in the case of
$\alpha_{i+ 1/2} \ne 0$ we have to use the principle of maximum
\cite{Samarsky1977}, that for the scheme (\ref{EQ_3002}) gives
the inequalities:

\begin{eqnarray}
1
-
\frac{1}{2} \alpha_{i+1/2}^{-}
+
\frac{1}{2}
\frac{\alpha_{i-1/2}^{-}}{R_{i-1/2}^{-}}
\ge
0\,, \nonumber
\qquad\qquad\qquad \\
1
-
\frac{1}{2} \alpha_{i-1/2}^{+}
+
\frac{1}{2}
\frac{\alpha_{i+1/2}^{+}}{R_{i+1/2}^{+}}
\ge
0\,,
\label{EQ_3003_5}
\qquad\qquad\qquad \\
\frac{w-v}{2h}\tau
\left(
1
-
\frac{1}{2} \alpha_{i+1/2}^{-}
+
\frac{1}{2}
\frac{\alpha_{i-1/2}^{-}}{R_{i-1/2}^{-}}
\right)
+
\frac{w+v}{2h}\tau
\left(
1
-
\frac{1}{2} \alpha_{i-1/2}^{+}
+
\frac{1}{2}
\frac{\alpha_{i+1/2}^{+}}{R_{i+1/2}^{+}}
\right)
\nonumber \\
\le
1\,,
\qquad\qquad\qquad \nonumber
\end{eqnarray}
where

\[
R_{i+1/2}^{+}
=
\frac{\rho_{i}^{n}-\rho_{i-1}^{n}}
{\rho_{i+1}^{n}-\rho_{i}^{n}}\,, \qquad\qquad
R_{i-1/2}^{-}
=
\frac{\rho_{i+1}^{n}-\rho_{i}^{n}}
{\rho_{i}^{n}-\rho_{i-1}^{n}}
\]
are smoothness analyzers. Note that last inequality
(\ref{EQ_3003_5}) determines also the stability condition for
the scheme (\ref{EQ_3002}).

To monotonize the scheme (\ref{EQ_3002}) a number of approaches
were suggested (see, e.g.,
\cite{VanLeer1974,VanLeer1977,Roe1985,Sweby1984}). Vyaznikov et
al. \cite{Vyaznikov1989} proposed to consider the coefficients
$\alpha$ as a piecewise-linear functions of the analyzers $R$.
These functions must satisfy to the following conditions:
1)~$\alpha = 0$, $R \le 0$; 2)~$0 \le \alpha \le 2$, $R \ge 0$.
Let us take $\alpha(R)$ in the form of two-parametric set of
piecewise-linear functions proposed by Chakravarthy and Osher
\cite{ChakravarthyOsher1985}:

\[
\alpha(R) =
\frac{1+\psi}{2}\cdot\beta\cdot\minmod(1/\beta,R) +
\frac{1-\psi}{2}\cdot\minmod(\beta,R)\,,
\label{EQ_3006}
\]
where

\[
\minmod(x,y) = \frac{1}{2} (\sign(x)+\sign(y))\min(|x|,|y|)\,.
\]
The function $\minmod(x,y)$ equals to zero if $x$ and $y$ have
the different signs and to argument with the minimal modulus
otherwise. The function $\alpha(R)$ is linear in the region of
the smooth solution: $1/\beta \le R \le \beta$,

\begin{equation}
\alpha(R)
=
\frac{1+\psi}{2}
+
\frac{1-\psi}{2}R
\end{equation}
and increases from value
$\alpha_{1}=((1+\psi)\beta+(1-\psi))/(2\beta)$ to
$\alpha_{2}=((1+\psi)+(1-\psi)\beta)/2$.

The scheme (\ref{EQ_3002}) has the second or third order of
approximation on space in the regions of smooth solution. This
result can be obtained by of expanding of the scheme
(\ref{EQ_3002}) to the Taylor series in the neighborhood of
point $M = (t = t^{n}, x = x_{i})$. The corresponding first
difference approximation for the scheme (\ref{EQ_3002}) is:

\[
\frac{\partial \rho}{\partial t}
+
v \frac{\partial \rho}{\partial x}
+
\frac{v}{4}
\bigl (
\frac{1}{3} - \psi
\bigr )
\frac{\partial^{3} \rho}{\partial x^{3}}
h^{2}
+
O(\tau+h^{3})\,,
\]
Therefore at the value $\psi = 1/3$ in the region of smooth
solution the scheme (\ref{EQ_3002}) attains the third order of
approximation in space. For the other values of $\psi$ the
scheme (\ref{EQ_3002}) has the second order of approximation.

The parameters $\alpha_{1}$ and $\alpha_{2}$ determine the range
of values of the function $\alpha(R)$ in the smoothness region
(where $R\sim1$). The optimal value of the parameter $\beta$
attains at $\alpha_{2}=2$ when
$\beta=\beta_{max}=(3-\psi)/(1-\psi)$. In this case we can find
that $\alpha_{1}=2/(3-\psi)$, $\alpha_{2}=2$. The stability
condition is determined by last inequality (\ref{EQ_3003_5}). It
can be rewritten in the following form:

\begin{eqnarray}
\qquad\qquad
\frac{w \tau}{h}
& +
& \frac{(w-v)\tau}{4h}
\max\limits_{R} \left\{ \frac{\alpha}{R} \right \}
-
\frac{(w-v)\tau}{4h}
\min\limits_{R} \left\{ \alpha \right \} \nonumber \\
~\nonumber\\
& +
& \frac{(w+v)\tau}{4h}
\max\limits_{R} \left\{ \frac{\alpha}{R} \right \}
-
\frac{(w+v)\tau}{4h}
\min\limits_{R} \left\{ \alpha \right \}
\le
1\,. \nonumber
\end{eqnarray}
From this inequality and from the definition of $\alpha(R)$ we
can obtain the stability condition of Courant, Friedrichs and
Lewy (CFL condition, \cite{CFL1928}):

\begin{equation}
\frac{w \tau}{h} \le \frac{4}{5-\psi+\beta (1+\psi)}\,.
\label{EQ_3007a}
\end{equation}
In the case of optimal values of the parameters $\psi = 1/3$ and
$\beta = 4$ we can rewrite (\ref{EQ_3007a}) by the following
way:

\[
\frac{w \tau}{h} \le 0.4\,.
\]
The developed scheme can be rewritten more briefly in the
conservative form:

\begin{equation}
\frac{\rho_{i}^{n+1} - \rho_{i}^{n}}{\tau}
+
\frac{f_{i+1/2} - f_{i-1/2}}{h}
=
0\,,
\label{EQ_3008}
\end{equation}
where

\begin{eqnarray}
\qquad\qquad
f_{i+1/2}
& =
& f_{i+1/2}^{0}
-
\frac{1-\psi}{4}
\minmod (f_{i+3/2}^{-}, \beta f_{i+1/2}^{-}) \nonumber \\
& -
& \frac{1+\psi}{4}
\minmod (f_{i+1/2}^{-}, \beta f_{i+3/2}^{-}) \nonumber \\
& +
& \frac{1+\psi}{4}
\minmod (f_{i+1/2}^{+}, \beta f_{i-1/2}^{+}) \nonumber \\
& +
& \frac{1-\psi}{4}
\minmod (f_{i-1/2}^{+}, \beta f_{i+1/2}^{+})\,,
\label{EQ_3009}
\end{eqnarray}

\[
f_{i+1/2}^{0}
=
\frac{f_{i+1}+f_{i}}{2}
-
\frac{w}{2}
\left (
\rho_{i+1}^{n}
-
\rho_{i}^{n}
\right ),
\]

\[
f_{i+1/2}^{-} = f_{i+1/2}^{0} - f_{i}\,, \qquad\qquad
f_{i+1/2}^{+} = f_{i+1} - f_{i+1/2}^{0}
\]
and fluxes $f_{i} = v\rho_{i}^{n}$.

\subsection{System of hyperbolic equations}

The developed methodology of difference scheme construction for
the linear advection equation can be generalized onto the system
of the linear hyperbolic equations:

\begin{equation}
\frac{\partial \vec{u}}{\partial t}
+
{\sf A} \frac{\partial \vec{u}}{\partial x}
=
0\,,
\label{EQ_3101}
\end{equation}
where $\vec{u}$ is the vector of the unknown variables and ${\sf
A}$ is the matrix of the constant coefficients (hyperbolicity
matrix). This system can be rewritten in the conservative form:

\[
\frac{\partial \vec{u}}{\partial t}
+
\frac{\partial \vec{F}}{\partial x}
=
0
\]
with the vector of fluxes: $\vec{F} = {\sf A} \vec{u}$.

The system of equations (\ref{EQ_3101}) can be represented in
the equivalent form of the equations for the Riemann invariants

\begin{equation}
S^{\alpha} = (\vec{l}^{\alpha} \cdot \vec{u})
\label{EQ_3103XXX}
\end{equation}
as

\begin{equation}
\frac{\partial S^{\alpha}}{\partial t}
+
\lambda_{\alpha}
\frac{\partial S^{\alpha}}{\partial x}
=
0\,,
\label{EQ_3103}
\end{equation}
where $\vec{l}^{\alpha}$ are the left eigenvectors of the matrix
${\sf A}$ and $\lambda_{\alpha}$ are its eigenvalues. The
original variables $\vec{u}$ can be obtained using the inverse
transformation:

\begin{equation}
\vec{u}
=
\sum\limits_{\alpha=1}^{N}
\vec{r}_{\alpha} S^{\alpha}\,,
\label{EQ_3104}
\end{equation}
where $\vec{r}_{\alpha}$ are the right eigenvectors of the matrix
${\sf A}$.

We approximate each equation in the system (\ref{EQ_3103}) by
the difference scheme (\ref{EQ_3008}):

\begin{equation}
\frac{S_{i}^{\alpha,n+1} - S_{i}^{\alpha,n}}{\tau}
+
\frac{H^{\alpha}_{i+1/2} - H^{\alpha}_{i-1/2}}{h}
=
0\,,
\label{EQ_3105}
\end{equation}
where fluxes

\begin{eqnarray}
\qquad\qquad
H^{\alpha}_{i+1/2}
& =
& H^{\alpha,0}_{i+1/2}
-
\frac{1-\psi}{4}
\minmod (H^{\alpha -}_{i+3/2}, \beta H^{\alpha -}_{i+1/2})
\nonumber \\
& -
& \frac{1+\psi}{4}
\minmod (H^{\alpha -}_{i+1/2}, \beta H^{\alpha -}_{i+3/2})
\nonumber \\
& +
& \frac{1+\psi}{4}
\minmod (H^{\alpha +}_{i+1/2}, \beta H^{\alpha +}_{i-1/2})
\nonumber \\
& +
& \frac{1-\psi}{4}
\minmod (H^{\alpha +}_{i-1/2}, \beta H^{\alpha +}_{i+1/2})\,,
\nonumber
\end{eqnarray}

\[
H^{\alpha,0}_{i+1/2}
=
\frac{H^{\alpha}_{i+1}+H^{\alpha}_{i}}{2}
-
\frac{w_{\alpha}}{2}
\left (
S^{\alpha,n}_{i+1}
-
S^{\alpha,n}_{i}
\right )\,, \ \
H^{\alpha}_{i} = \lambda_{\alpha} S^{\alpha,n}_{i}\,, \ \
w_{\alpha} \ge |\lambda_{\alpha}|\,,
\]

\[
H^{\alpha -}_{i+1/2} = H^{\alpha,0}_{i+1/2} - H^{\alpha}_{i}\,, \ \
H^{\alpha +}_{i+1/2} = H^{\alpha}_{i+1} - H^{\alpha,0}_{i+1/2}\,.
\]

To construct the scheme for the variables $\vec{u}$ we have to
perform the transformation (\ref{EQ_3103XXX}) in
(\ref{EQ_3105}). Finally the scheme reads:

\begin{equation}
\frac{\vec{u}^{n+1}_{i} - \vec{u}^{n}_{i}}{\tau}
+
\frac{\vec{F}_{i+1/2} - \vec{F}_{i-1/2}}{h}
=
0\,,
\label{EQ_3109}
\end{equation}
where fluxes are:

\begin{eqnarray}
\qquad\qquad
\vec{F}_{i+1/2}
& =
& \vec{F}^{0}_{i+1/2}
-
\frac{1-\psi}{4}
\sum\limits_{\alpha}
\minmod (\vec{F}^{\alpha -}_{i+3/2}, \beta \vec{F}^{\alpha -}_{i+1/2})
\nonumber \\
& -
& \frac{1+\psi}{4}
\sum\limits_{\alpha}
\minmod (\vec{F}^{\alpha -}_{i+1/2}, \beta \vec{F}^{\alpha -}_{i+3/2})
\nonumber \\
& +
& \frac{1+\psi}{4}
\sum\limits_{\alpha}
\minmod (\vec{F}^{\alpha +}_{i+1/2}, \beta \vec{F}^{\alpha +}_{i-1/2})
\nonumber \\
& +
& \frac{1-\psi}{4}
\sum\limits_{\alpha}
\minmod (\vec{F}^{\alpha +}_{i-1/2}, \beta \vec{F}^{\alpha +}_{i+1/2})\,,
\label{EQ_3110NEW}
\end{eqnarray}

\[
\vec{F}^{0}_{i+1/2}
=
\frac{\vec{F}_{i+1}+\vec{F}_{i}}{2}
-
\frac{1}{2}
{\sf R} {\sf W} {\sf L}
\left (
\vec{u}^{n}_{i+1}
-
\vec{u}^{n}_{i}
\right )\,,
\]

\[
{\sf W}
=
{\diag} (w_{1}, w_{2}, \ldots, w_{N})\,,
\]

\[
\vec{F}^{\alpha \pm}_{i+1/2}
=
\frac{1}{2}
\left ( \lambda_{\alpha} \pm w_{\alpha} \right )
\left (
\vec{l}_{\alpha}
\cdot
\left (
\vec{u}^{n}_{i+1} - \vec{u}^{n}_{i}
\right )
\right )
\vec{r}^{\alpha}\,,
\]
where ${\sf L}$, ${\sf R}$ -- matrixes of the left and of the
right eigenvectors of the matrix ${\sf A}$ respectively.

The difference scheme (\ref{EQ_3109},\ref{EQ_3110NEW}) satisfies
to the total variation diminishing (TVD) principle
\cite{Harten1984}, since it is constructed on the basis of the
monotonic schemes for the Riemann invariants. In the present
paper we consider the Lax--Friedrichs schemes with the
coefficients
$w_{\alpha}=w\ge\max\limits_{\alpha}\{|\lambda_{\alpha}|\}$.
For this case the fluxes in the base scheme reads
\cite{Lax1954,Lax1957}:

\begin{equation}
\vec{F}^{0}_{i+1/2}
=
\frac{\vec{F}_{i+1}+\vec{F}_{i}}{2}
-
\frac{w}{2}
\left (
\vec{u}^{n}_{i+1}
-
\vec{u}^{n}_{i}
\right )\,.
\label{EQ_3114}
\end{equation}

During the calculation of fluxes (\ref{EQ_3110NEW}) the
numerical procedure of the $\minmod$-evaluation should be called
very often. To decrease the computation time we consider a more
simple scheme in which the additional fluxes are summed up.
Finally the scheme (\ref{EQ_3109}) is determined by the fluxes:

\begin{eqnarray}
\qquad\qquad
\vec{F}_{i+1/2}
& =
& \vec{F}^{0}_{i+1/2}
-
\frac{1-\psi}{4}
\minmod (\vec{F}^{-}_{i+3/2}, \beta \vec{F}^{-}_{i+1/2})
\nonumber \\
& -
& \frac{1+\psi}{4}
\minmod (\vec{F}^{-}_{i+1/2}, \beta \vec{F}^{-}_{i+3/2})
\nonumber \\
& +
& \frac{1+\psi}{4}
\minmod (\vec{F}^{+}_{i+1/2}, \beta \vec{F}^{+}_{i-1/2})
\nonumber \\
& +
& \frac{1-\psi}{4}
\minmod (\vec{F}^{+}_{i-1/2}, \beta \vec{F}^{+}_{i+1/2})\,,
\label{EQ_3115}
\end{eqnarray}
where

\[
\vec{F}^{-}_{i+1/2}
=
\vec{F}^{0}_{i+1/2}
-
\vec{F}_{i}, \qquad \qquad
\vec{F}^{+}_{i+1/2}
=
\vec{F}_{i+1}
-
\vec{F}^{0}_{i+1/2}
\label{EQ_3116}
\]
and the fluxes $\vec{F}^{0}_{i+1/2}$ are determined by the
expression (\ref{EQ_3114}).

This methodology can be generalized easy onto the system of the
nonlinear equations:

\begin{equation}
\frac{\partial \vec{u}}{\partial t}
+
\frac{\partial \vec{F}(\vec{u})}{\partial x}
=
0\,.
\label{EQ_3117}
\end{equation}

The finite-difference scheme (\ref{EQ_3109}) for the system
(\ref{EQ_3117}) is determined by the same fluxes
(\ref{EQ_3114},\ref{EQ_3115}), but the coefficients $w$ we
should take in the form:

\begin{equation}
w
=
w_{i+1/2}
=
\phi
\max\limits_{\alpha}
\left \{
|\lambda_{\alpha,i}|, |\lambda_{\alpha,i+1}|
\right \}\,, \ \
\phi \ge 1\,,
\label{EQ_3118}
\end{equation}
where $\lambda_{\alpha}$ are the eigenvalues of the
hyperbolicity matrix ${\sf A}=\partial \vec{F}/\partial
\vec{u}$.

\subsection{MHD modification of the finite-difference scheme}

We construct now the finite-difference TVD scheme for the
isothermal or adiabatical 1D MHD in Cartesian
coordinates. We assume that the magnetic field and the velocity
have all three components: $\vec{B} = \{ B_{x}, B_{y}, B_{z}
\}$, $\vec{v} = \{ v_{x}, v_{y}, v_{z} \}$.

The MHD equations in the conservative form,

\[
\frac{\partial \vec{u}}{\partial t}
+
\frac{\partial \vec{F}}{\partial x}
=
0
\]
are determined by the vector of the conservative variables:

\[
\vec{u}
=
\left\{
\rho\,, \
\rho v_{x}\,, \ \rho v_{y}\,, \ \rho v_{z}\,, \
B_{x}\,, \ B_{y}\,, \ B_{z}\,, \
\rho \varepsilon
+
\rho \frac{\vec{v}^{2}}{2}
+
\frac{\vec{B}^{2}}{8 \pi}
\right\}^{T}
\]
and by the vector of the fluxes:

\begin{eqnarray}
{\vec{F}}
& =
& \left\{
\rho v_{x}\,, \
P + \rho v_{x}^{2}
+
\frac{\vec{B}^{2}}{8 \pi} - \frac{B_{x}^{2}}{4 \pi}\,, \
\rho v_{x} v_{y} - \frac{B_{x}B_{y}}{4 \pi}\,, \
\rho v_{x} v_{z} - \frac{B_{x}B_{z}}{4 \pi}\,, \
\right. \nonumber \\
&&\qquad\qquad
0\,, \
v_{x}B_{y} - v_{y}B_{x}\,, \
v_{x}B_{z} - v_{z}B_{x}\,, \
\nonumber \\
&&\qquad\qquad
\left.
\rho v_{x}
\left(
\varepsilon
+
\frac{P}{\rho}
+
\frac{\vec{v}^{2}}{2}
+
\frac{\vec{B}^{2}}{4 \pi}
\right)
-
\frac{B_{x}}{4 \pi}
\left(
v_{x}B_{x} + v_{y}B_{y} + v_{z}B_{z}
\right)
\right\}^{T}\,.
\nonumber
\end{eqnarray}
For the adiabatic case the hyperbolicity matrix ${\sf A}$ is

\begin{equation}
\left (
\begin{array}{cccccccc}
0 & 1 & 0 & 0 & 0 & 0 & 0 & 0 \\
~\\
A_{21} & (3-\gamma)v_{x} & (1-\gamma)v_{y} & (1-\gamma)v_{z} & %%@
-\gamma\dfrac{B_{x}}{4\pi} & A_{26} & A_{27} & \gamma-1 \\
~\\
-v_{x}v_{y} & v_{y} & v_{x} & 0 & -\dfrac{B_{y}}{4\pi} & -\dfrac{B_{x}}{4\pi} & 0 %%@
& 0 \\
~\\
-v_{x}v_{z} & v_{z} & 0 & v_{x} & -\dfrac{B_{z}}{4\pi} & 0 & -\dfrac{B_{x}}{4\pi} %%@
& 0 \\
~\\
0 & 0 & 0 & 0 & 0 & 0 & 0 & 0 \\
~\\
A_{61} & \dfrac{B_{y}}{\rho} & -\dfrac{B_{x}}{\rho} & 0 & -v_{y} & v_{x} & 0 & 0 %%@
\\
~\\
A_{71} & \dfrac{B_{z}}{\rho} & 0 & -\dfrac{B_{x}}{\rho} & -v_{z} & 0 & v_{x} & 0 %%@
\\
~\\
A_{81} & A_{82} & A_{83} & A_{84} & A_{85} & A_{86} & A_{87} & \gamma v_{x}
\end{array}
\right )
\label{EQ_aaaa}
\end{equation}
where:

\[
A_{21} = \frac{\gamma-1}{2}\vec{v}^{2}-v_{x}^{2}\,, \qquad
A_{26} = (2-\gamma)\frac{B_{y}}{4\pi}\,, \qquad
A_{27} = (2-\gamma)\frac{B_{z}}{4\pi}\,,
\]
\[
A_{61} = \frac{v_{y}B_{x}-v_{x}B_{y}}{\rho}\,, \qquad
A_{71} = \frac{v_{z}B_{x}-v_{x}B_{z}}{\rho}\,,
\]
\[
A_{81} = -v_{x}
\left(
\frac{c^{2}}{\gamma-1}
+
\left(1-\frac{\gamma}{2}\right)\vec{v}^{2}
+
\vec{a}^{2}
\right)
+
a_{x} (\vec{a}\cdot\vec{v})\,,
\]
\[
A_{82} = \frac{c^{2}}{\gamma-1}
+
\frac{\vec{v}^{2}}{2}
+
\vec{a}^{2}
-
a_{x}^{2}
-
(\gamma-1)v_{x}^{2}\,,
\]
\[
A_{83} = (1-\gamma)v_{x}v_{y} - a_{x}a_{y}\,, \ \
A_{84} = (1-\gamma)v_{x}v_{z} - a_{x}a_{z}\,,
\]
\[
A_{85} = -\gamma v_{x}\frac{B_{x}}{4\pi}
-
v_{y}\frac{B_{y}}{4\pi}
-
v_{z}\frac{B_{z}}{4\pi}\,,
\qquad
A_{86} = -\gamma v_{x}\frac{B_{y}}{4\pi}
+
v_{x}\frac{B_{y}}{2\pi}
-
v_{y}\frac{B_{x}}{2\pi}\,,
\]
\[
A_{87} = -\gamma v_{x}\frac{B_{z}}{4\pi}
+
v_{x}\frac{B_{z}}{2\pi}
-
v_{z}\frac{B_{x}}{2\pi}\,.
\]
We use here the sound speed, $c=\sqrt{\gamma P/\rho}$ and
$\vec{a}=\vec{B}/\sqrt{4 \pi \rho}$. The calculations give the
following expressions for the eigenvalues of the hyperbolicity
matrix (\ref{EQ_aaaa}):

\[
\lambda_{1} = 0\,, \ \ \lambda_{2} = v_{x}\,, \ \
\lambda_{3,4} = v_{x} \pm a_{x}\,,
\]
\begin{equation}
\lambda_{5,6,7,8}
=
v_{x}
\pm
\sqrt
{
\frac{c^{2}+\vec{a}^{2}}{2}
\pm
\frac{1}{2}\sqrt
{
(c^{2}+\vec{a}^{2})^{2}
-
4 a_{x}^{2} c^{2}
}
}.
\label{EQ_3200}
\end{equation}
The zero value of $\lambda_{1}$ is the consequence of zero value
of the flux $\vec{F}$ component corresponding to $B_{x}$. The
value of $\lambda_{2}$ corresponds to the entropic wave,
$\lambda_{3,4}$ correspond to the Alfv\'en waves, and
$\lambda_{5,6,7,8}$ correspond to the fast and slow magnetosonic
waves. In the isothermal case the entropic wave disappears and
instead adiabatical sound speed it should be used the isothermal
sound speed $c_{T}$.

We should construct now the calculation rule for the
coefficients $w$ in (\ref{EQ_3118}). The values of these
coefficients should be not less than the maximal modulus of all
eigenvalues of the hyperbolicity matrix. It is clear that this
condition can be satisfied by the following way:

\begin{equation}
w_{i+1/2}
=
\phi
\max
\{
|v_{x,i}| + \Psi_{x,i}\,,
|v_{x,i+1}| + \Psi_{x,i+1}
\}\,, \ \
\phi \ge 1\,,
\label{EQ_3201}
\end{equation}
where

\[
\Psi_{x}
=
\sqrt
{
\frac{c^{2}+\vec{a}^{2}}{2}
+
\frac{1}{2} \sqrt
{
(c^{2}+\vec{a}^{2})^{2}
-
4 a_{x}^{2} c^{2}
}
}\,.
\]
Therefore we can keep in the relation (\ref{EQ_3118}) only
eigenvalues corresponding to the fast MHD waves. This allow us
to adapt the scheme
(\ref{EQ_3109},\ref{EQ_3114},\ref{EQ_3115},\ref{EQ_3201}) to the
simulation of the 1D MHD flows. If the magnetic field is absent
the fast magnetosonic speed $\Psi_{x}$ equals the sound speed
$c$, therefore the transition to the gasdynamics occurs
correctly. With the help of the splitting technique we can
construct the scheme for multidimensional (2D and 3D) MHD flows.
One of this scheme for the axisymmetric case we will describe in
the next section.

\section{Numerical method}

\subsection{General scheme}

The computational grid is determined by points $(r_{i}, z_{j})$
indexed by integer indexes $i=0, 1, \ldots, I$; $j=0,1, \ldots,
J$. These points denote the centers of the grid cells. The
coordinates of the cell boundaries are defined as the arithmetic
mean of corresponding coordinates of the cell centers and have
half-integer indexes. The computational variables are related
to the centers of the cells and the numerical fluxes are related
to the cell boundaries. Above we developed the TVD scheme in the
frame of 1D MHD approximation. Now we generalize this approach
onto the 2D MHD approximation. The `base' difference scheme for
equations (\ref{EQ_2008}) is the following one:

\[
\frac
{\vec{u}_{i,j}^{n+1} - \vec{u}_{i,j}^{n}}
{\triangle t}
+
\frac
{\vec{F}_{i+1/2,j} - \vec{F}_{i-1/2,j}}
{\triangle r}
+
\frac
{\vec{G}_{i,j+1/2} - \vec{G}_{i,j-1/2}}
{\triangle z}
=
\vec{R}_{i,j}\,.
\]
The main problem here is the definition of the calculation rule
for the numerical fluxes across the cell boundaries.

Let us consider for simplicity only the radial direction because
of the expressions of fluxes for the vertical direction can be
derived by similar way. The radial fluxes $\vec{F}_{i+1/2,j}$
and $\vec{F}_{i-1/2,j}$ should be calculated with the help of
the formulae (\ref{EQ_3114},\ref{EQ_3115},\ref{EQ_3118}).
Relatively determination of the coefficients $w$ we should note
the following.

In the definition of the vectors $\vec{u}$, $\vec{F}$ and
$\vec{G}$ in the form (\ref{EQ_2009A}--\ref{EQ_2009B}) the
radial coordinate $r$ is present. Therefore to calculate
eigenvalues of the hyperbolicity matrices we can close formally
the system (\ref{EQ_2008}) by the following obvious equation:

\[
\frac{\partial r}{\partial t} = 0\,.
\]
It can be shown that this extension of the original MHD system
not changes the standard set of the eigenvalues.

In fact we consider the following system of MHD equations:

\begin{equation}
\frac{\partial \vec{w}}{\partial t}
+
{\sf A}
\frac{\partial \vec{w}}{\partial r}
+
{\sf B}
\frac{\partial \vec{w}}{\partial z}
=
\vec{R}\,,
\label{EQ_4102}
\end{equation}
with the vector $\vec{w}$ determined as

\[
\vec{w}
=
\left\{
r\,, \ \rho\,, \
\rho v_{r}\,, \ \rho v_{\varphi}\,, \ \rho v_{z}\,, \
B_{r}\,, \ B_{\varphi}\,, \ B_{z}\,, \
\rho\left(\varepsilon
+
\frac{\vec{v}^{2}}{2}\right)
+
\frac{\vec{B}^{2}}{8 \pi}
\right\}^{T}\,.
\]

Obviously in Cartesian coordinates the eigenvalues of matrices
${\sf A}$ and ${\sf B}$ are determined by the same expressions
(\ref{EQ_3200}). Let us consider another vector of the variables
$\vec{u}$ that consists of $r$ and eight components of the
vector (\ref{EQ_2009A}). Making the transformation of the
equation (\ref{EQ_4102}) to the new variables $\vec{u}$ we
obtain:

\[
\frac{\partial \vec{u}}{\partial t}
+
{\sf A}'
\frac{\partial \vec{u}}{\partial r}
+
{\sf B}'
\frac{\partial \vec{u}}{\partial z}
=
\vec{R}'\,,
\]
where

\begin{equation}
{\sf A}' = {\sf T}^{-1} {\sf A} {\sf T}\,, \qquad
{\sf B}' = {\sf T}^{-1} {\sf B} {\sf T}\,, \qquad
\vec{R}' = {\sf T}^{-1} \vec{R}\,,
\label{EQ_4103}
\end{equation}
and matrix ${\sf T}=\partial \vec{u}/\partial \vec{w}$ is
Jacobian of transformation. The determinant of this matrix $\det
\hat{\sf T} = 1/r^{7}$ is not equal to zero. Hence the formulae
(\ref{EQ_4103}) describes the similarity transformations of
matrices ${\sf A}$ and ${\sf B}$ that does not change their
eigenvalues.

This discussion allows us to take the values of the coefficients
$w_{i+1/2}$ in (\ref{EQ_3118}) in the following form:

\begin{equation}
w_{i+1/2}
=
\phi
\max
\{
|v_{r,i}| + \Psi_{r,i}\,,
|v_{r,i+1}| + \Psi_{r,i+1}
\},\, \qquad
\phi \ge 1\,,
\label{EQ_4104}
\end{equation}

where
\begin{equation}
\Psi_{r}
=
\sqrt
{
\frac{c^{2}+\vec{a}^{2}
+
\sqrt
{
(c^{2}+\vec{a}^{2})^{2}
-
4 a_{r}^{2} c^{2}
}
}{2}
}\,.
\label{EQ_4105}
\end{equation}
These coefficients are determined by eigenvalues that correspond
to the fast magnetosonic wave. The expressions (\ref{EQ_4104})
can be applied both to the adiabatical and to the isothermal
plasma flows. The energy equation should be excluded in last
case and as $c$ in (\ref{EQ_4105}) should be used the isothermal
sound speed $c_{T}$.

Since the fluxes related to the different space directions are
constructed independently, the $z$-part of the numerical scheme
can be developed by similar way.

\subsection{Poisson solver}

Poisson equation for the gravitational potential is solved by
ADI-method (alternating directions implicit method) of Douglas
and Rachford
\cite{PeacemanRachford1955,Douglas1955,DouglasRachford1956}:

\[
\frac{\bar{\Phi}-\Phi^{p}}{\tau_{p}}
=
\theta \Lambda_{rr} \bar{\Phi}
+
(1-\theta) \Lambda_{zz} \Phi^{p}
-
\frac{\rho}{2}
\]
\[
\frac{\Phi^{p+1} - \bar{\Phi}}{\tau_{p}}
=
(1-\theta) \Lambda_{rr} \bar{\Phi}
+
\theta \Lambda_{zz} \Phi^{p+1}
-
\frac{\rho}{2}
\]
where $\theta$ is the parameter of scheme implicity,
$\Lambda_{rr}$, $\Lambda_{zz}$ are second order
finite-difference approximations of the corresponding parts of
Laplacian operator in $r$-- and $z$--directions, respectively,
$\tau_{p}$ is the iteration parameter.

Boundary conditions for the gravitational potential on the inner
boundaries are regular: $\partial\Phi/\partial\vec{n}=0$.
Boundary condition on the external boundary is
$\Phi_{int}=\Phi_{ext}$ where $\Phi_{int}$ is the computing
potential, $\Phi_{ext}$ is the external potential of the
protostellar cloud. It can be obtained from Poisson integral
transformation of equation (\ref{EQ_2004}) solution:

\[
\Phi_{ext}(r,z)
=
- G \sum \limits_{l=0}^{\infty}
\frac{P_{l}(x)}{(r^{2}+z^{2})^{l+1}/2}
\int \limits_{V'} {\d}V' \rho(r',z')(r'^{2}+z'^{2})^{l/2}
P_{l}(x')\,,
\]
where $x=z(r^{2}+z^{2})^{-l/2}$, and $P_{l}(x)$ is the Legendre
polynomial of $l$th order. All terms with odd $l$ are absent in
this representation as a result of the axial symmetry.

\subsection{Implementation of the $\diver \vec{B} = 0$ condition}

Another difficulty of the numerical simulation of the
multidimensional MHD flows is connected with the necessity of
the numerical implementation of the vanishing divergence
condition for the magnetic field

\begin{equation}
\diver \vec{B} = 0\,.
\label{EQ_4301}
\end{equation}

Numerical errors due to discretization and finite-difference
differential operator splitting can lead to nonzero divergence
over the time. This problem can be solved by many ways
\cite{Toth2000}. In the first approach the 8-wave formulation
of the MHD equations is used (see also \cite{Powell1994}).
Another way uses the constrained transport suggested by Evans
and Hawley \cite{EvansHawley1988}. Finally, the third approach
is the projection scheme \cite{BrackbillBarnes1980}. Our
numerical scheme uses the last method (see
\cite{RyuJones1995a,Balsara1998b}).

Denote as $\vec{B}^{*}$ the magnetic field obtained by the
numerical solution of the induction equation (\ref{EQ_2002}).
This `numerical' magnetic field contains the numerical error
$\vec{b}$, that consist of vortex and potential parts:

\[
\vec{b} = [\nabla , \vec{a}] + \nabla \phi\,.
\]
Therefore exact solution of induction equation is:

\begin{equation}
\vec{B} = \vec{B}^{*} - \vec{b}\,.
\label{EQ_4302}
\end{equation}
Vortex part of numerical error $[\nabla , \vec{a}]$ does not
contributes to divergence of the magnetic field $\vec{B}$.
Substitution of (\ref{EQ_4302}) into (\ref{EQ_4301}) shows that
potential $\phi$ must satisfy the Poisson equation

\begin{equation}
\nabla^{2} \phi = \diver \vec{B}^{*}\,.
\label{EQ_4303XXX}
\end{equation}
Resolving this equation numerically for the defined `numerical'
magnetic field $\vec{B}^{*}$ for each time step we calculate the
divergence free magnetic field with the help of formula

\[
\vec{B}^{s} = \vec{B}^{*} - \nabla \phi\,.
\]
Further we will assume that $\vec{B} = \vec{B}^{s}$. The
algorithm is implemented in our numerical code as follows.
Denote the central cell with indexes ($i$, $j$) by the index $C$
for brevity (see. Fig.~1), the neighbor cells by indexes $W$,
$N$, $E$, $S$ and intermediate coordinates (cell boundaries) by
double indexes $CE$, $CS$ etc.

\begin{figure}
\centerline{\hbox{\psfig{figure=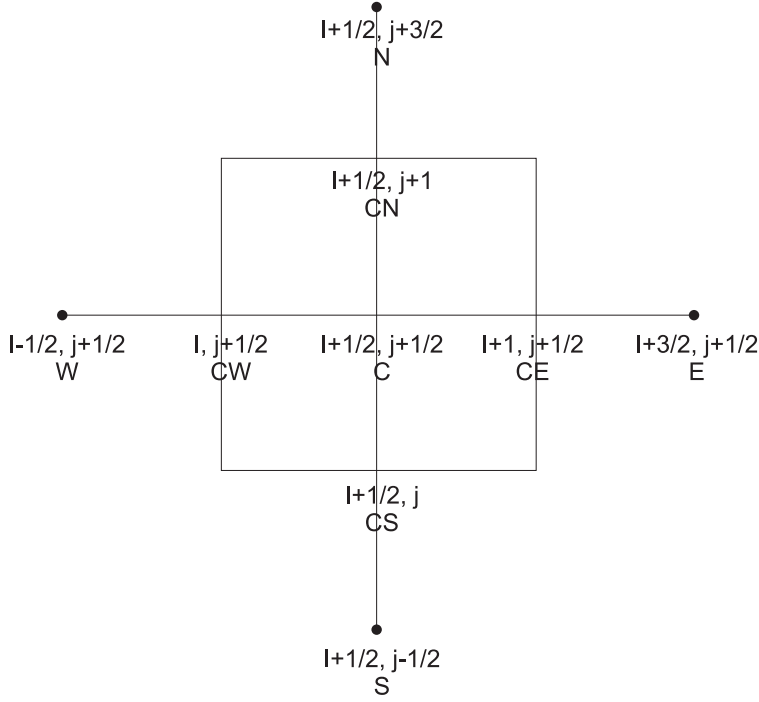,width=10cm}}}
\caption{A typical grid cell}
\end{figure}

Then we can approximate the divergence of magnetic field by the
expression:

\begin{equation}
\bigl ( \diver{\vec{B}} \bigr )_{C}
=
\frac{1}{r_{C}}
\frac{r_{CE}B_{r,CE} - r_{CW}B_{r,CW}}{r_{CE}-r_{CW}}
+
\frac{B_{z,CN} - B_{z,CS}}{z_{CN}-z_{CS}}\,,
\label{EQ_4303}
\end{equation}
where intermediate values are calculated by the formulae

\[
B_{r,CE}
=
\frac{1}{2}
\bigl (
B_{r,C} + B_{r,E}
\bigr )\,, \qquad
B_{r,CS}
=
\frac{1}{2}
\bigl (
B_{r,C} + B_{r,S}
\bigr )\,, \qquad
\mathrm{etc.}
\]
The finite-difference approximation of the Laplacian operator is

\begin{eqnarray}
\qquad\qquad
\bigl ( \triangle \phi \bigr )_{C}
& =
& \frac{1}{r_{C} (r_{CE} - r_{CW})}
\biggl (
r_{CE} \frac{\phi_{E}-\phi_{C}}{r_{E}-r_{C}}
-
r_{CW} \frac{\phi_{C}-\phi_{W}}{r_{C}-r_{W}}
\biggr ) \nonumber \\
& +
& \frac{1}{(z_{CN} - r_{CS})}
\biggl (
\frac{\phi_{N}-\phi_{C}}{z_{N}-z_{C}}
-
\frac{\phi_{C}-\phi_{S}}{z_{C}-z_{S}}
\biggr )\,. \nonumber
\end{eqnarray}

It is easy to see that if the potential $\phi$ is the exact
solution of the finite-difference Poisson equation
(\ref{EQ_4303XXX}) then the magnetic field $\vec{B}$ (for which
the finite-difference divergence (\ref{EQ_4303}) is equal to
zero exactly) can be obtained by the formulae:

\[
B_{r,C}
=
B^{*}_{r,C}
-
\frac{1}{2}
\biggl [
\frac{\phi_{E}-\phi_{C}}{r_{E}-r_{C}}
+
\frac{\phi_{C}-\phi_{W}}{r_{C}-r_{W}}
\biggr ]\,,
\]
\[
B_{z,C}
=
B^{*}_{z,C}
-
\frac{1}{2}
\biggl [
\frac{\phi_{N}-\phi_{C}}{z_{N}-z_{C}}
+
\frac{\phi_{C}-\phi_{S}}{z_{C}-z_{S}}
\biggr ]\,.
\]

The Poisson equation (\ref{EQ_4303XXX}) for the potential $\phi$
is solved in our scheme by the ADI-method that is described in
the previous section. As the additional boundary
conditions on the inner boundaries ($r=0$, $z=0$) we use the
relations $\partial \phi/\partial\vec{n} = 0$. On the outer
boundaries we use the boundary condition $\phi = 0$ since
the external field has identically vanishing divergence.

\section{Test computations}

\subsection{Background}

The numerical code `Moon' was developed on the basis of the
proposed finite-difference scheme. It can simulate the 1D and
2D selfgravitating MHD flows. The code is implemented
by the programming language C++ and completely object-oriented.
Source text of the program uses only standard C++ elements and
does not relate to concrete compiler. The kernel of the code
uses intensively the object-oriented library `Numerical Tool
Box' (NTB 3.5). This library was developed by A.G.Zhilkin in
1996 for more easy creation of the complex numerical codes. To
check the properties of our code we tested it on the number of
test problems with known (exact or approximate) analytical
solutions.

\subsection{One-dimensional tests}

\subsubsection{Test 1. Linear advection}

The linear advection of the square density profile was simulated
to check the phase-error properties of the numerical scheme.
Figure~2 gives the density profile after 100th time step without
(triangles) and with flux correction (dots). The computational
grid consists of 100 cells. The analytical solution is drawn by
the solid line.

\begin{figure}
\centerline{\hbox{\psfig{figure=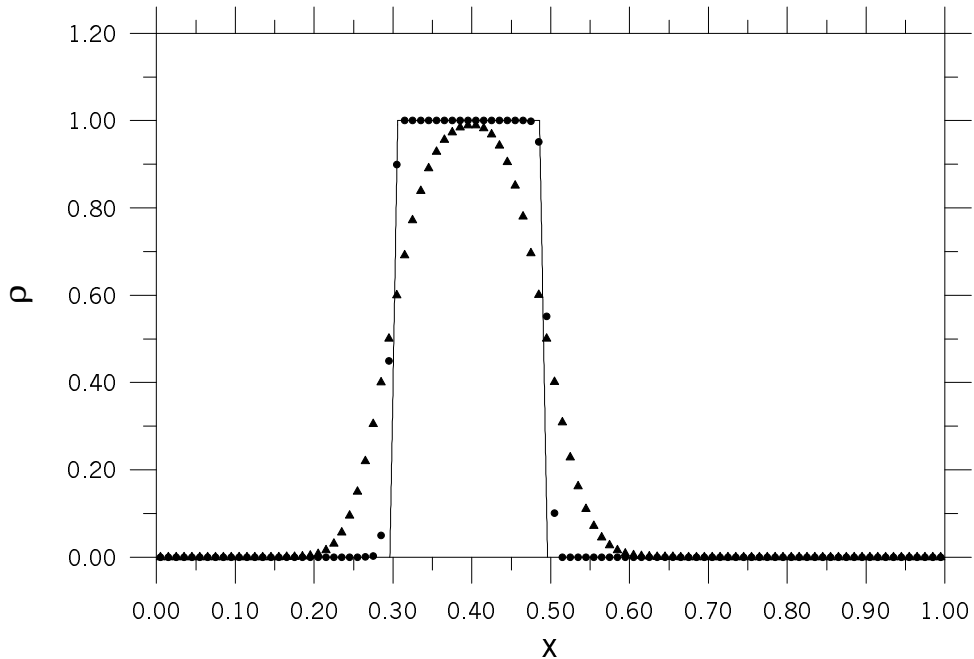,width=13cm}}}
\caption{Linear advection of squared density profile. The
density distributions is shown after 100 time-step computed by
the scheme of the 1st-order of approximation (triangles) and by
the scheme of the 3rd-order of approximation (dots). The solid
line corresponds to the analytical solution}
\end{figure}

It is seen that the 1st-order of approximation method has a
large numerical diffusion. This diffusion smears the initial
profile rapidly, the width of smearing being increased during
calculation. The method of the 3rd-order of approximation
produces a profile that is smeared onto 3--4 cells only and
conserves its form during a long time. We can say therefore
that amplitude and phase errors in the scheme of the 3rd-order
of approximation are selfconsistent and its order of
approximation is not less actually then the approximation order
of the so-called LPE-schemes (little phase error) (see
\cite{BorisBook1976}). In our opinion this circumstance is
connected with the successful selection of the antidiffusion
limiters (\ref{EQ_3009}) in the scheme (\ref{EQ_3008}).

\subsubsection{Test 2. Decay of arbitrary gasdynamical
discontinuity}

\begin{figure}
\centerline{\hbox{\psfig{figure=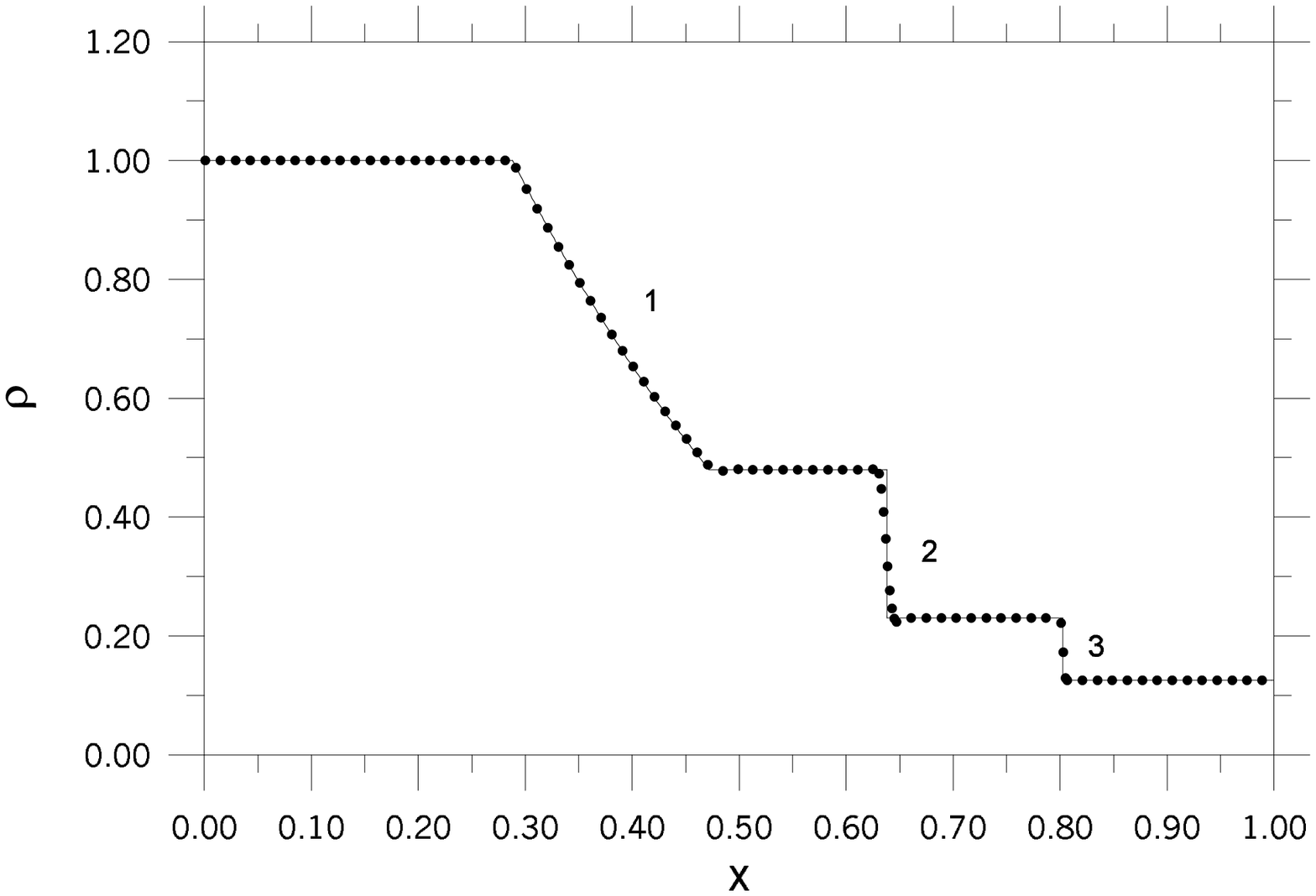,width=13cm}}}
\caption{Decay of an gasdynamical discontinuity. The
density distribution (dots) is shown at the time moment
$t=0.164$. The solid line corresponds to the analytical
solution. The digits correspond to: 1 --- rarefaction wave, 2
--- contact discontinuity, 3 --- shock wave. In the regions of
smooth flow the dots is displayed more seldom (each 7th point).
In the regions of contact discontinuity and shock wave all
points of numerical solution are displayed}
\end{figure}

Numerical simulation of the arbitrary discontinuity decay
(Riemann problem) allows us to check the scheme properties
relating to the resolution of the shock waves, contact
discontinuities and rarefaction waves. At the initial time
moment the computational domain along the $x$-axis was divided
on two subdomains $A$ and $B$. In the initial state the pressure
and density have the following values: in subdomain $A$ ---
$\rho_A=1$, $P_A=1$; in subdomain $B$ --- $\rho_B=0.125$,
$P_B=0.1$. The velocity in both domains $v_A=v_B=0$ and
adiabatic index $\gamma=5/3$. Figure~3 shows the analytical
solution (solid line) and the results of the numerical
computation (dots) at the time moment $t = 0.164$. As a result
of the discontinuity decay the rarefaction wave 1 propagates to
the left, contact discontinuity 2 and shock wave 3 propagates to
the right. The Figure shows that the scheme resolves the shock
wave sufficiently well, but gives the small non-physical
oscillations on the contact discontinuity. To smooth it is
necessary to use additional limiters of antidiffusion fluxes.

\begin{figure}[t]
\centerline{\hbox{\psfig{figure=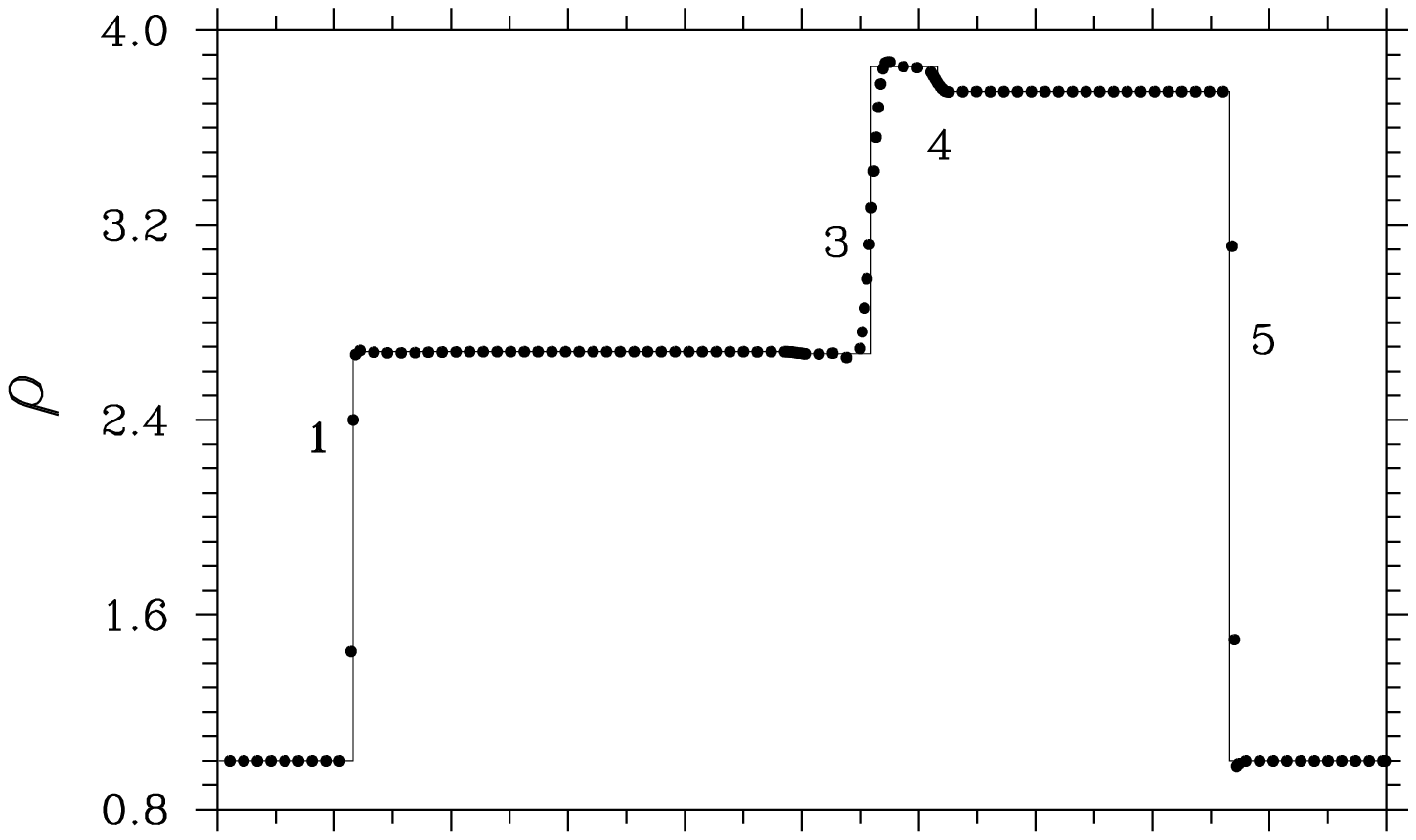,width=7cm}}
~~\hbox{\psfig{figure=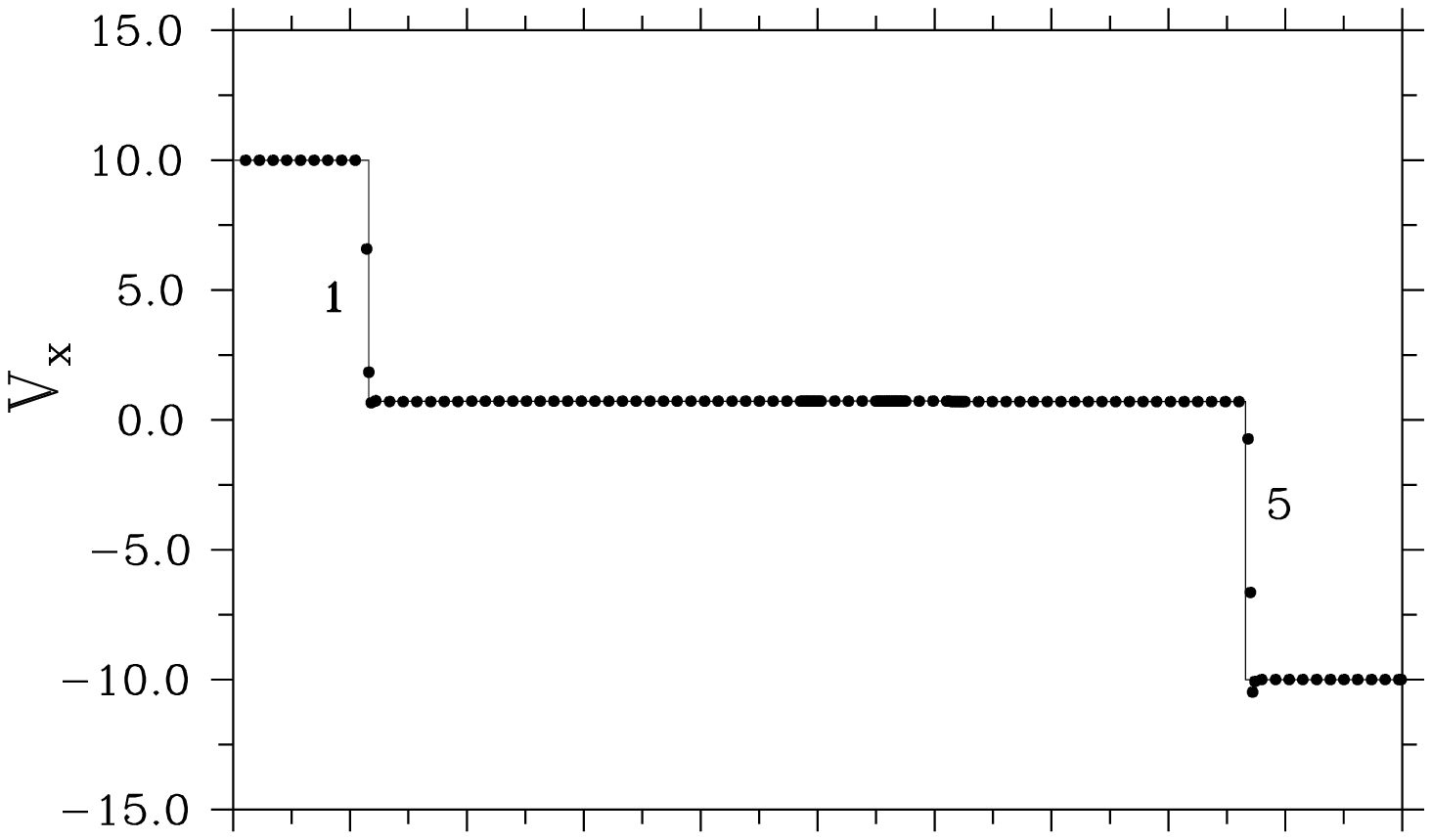,width=7cm}}}
\vspace{4mm}
\centerline{\hbox{\psfig{figure=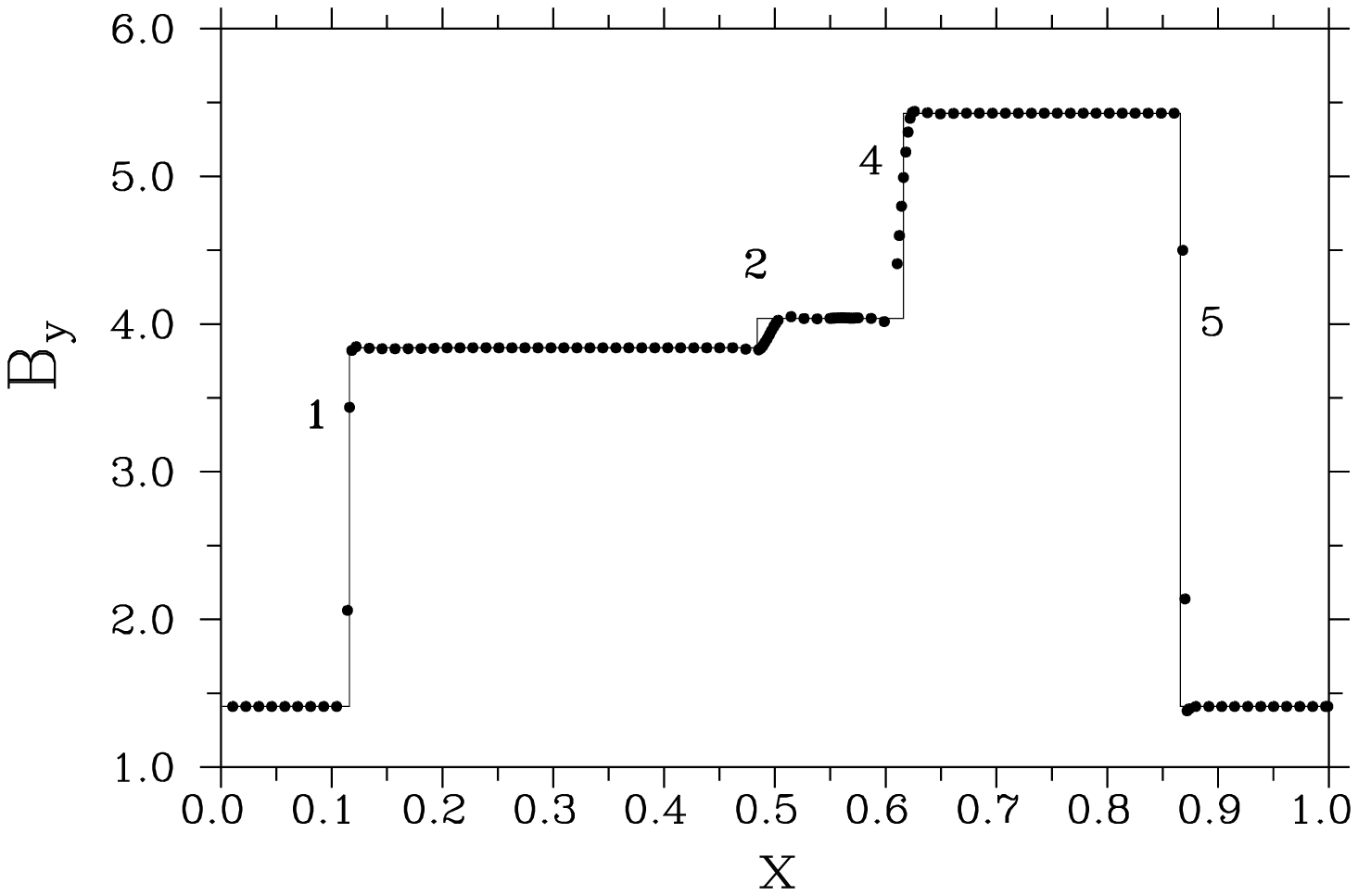,width=7cm}}
\hbox{\psfig{figure=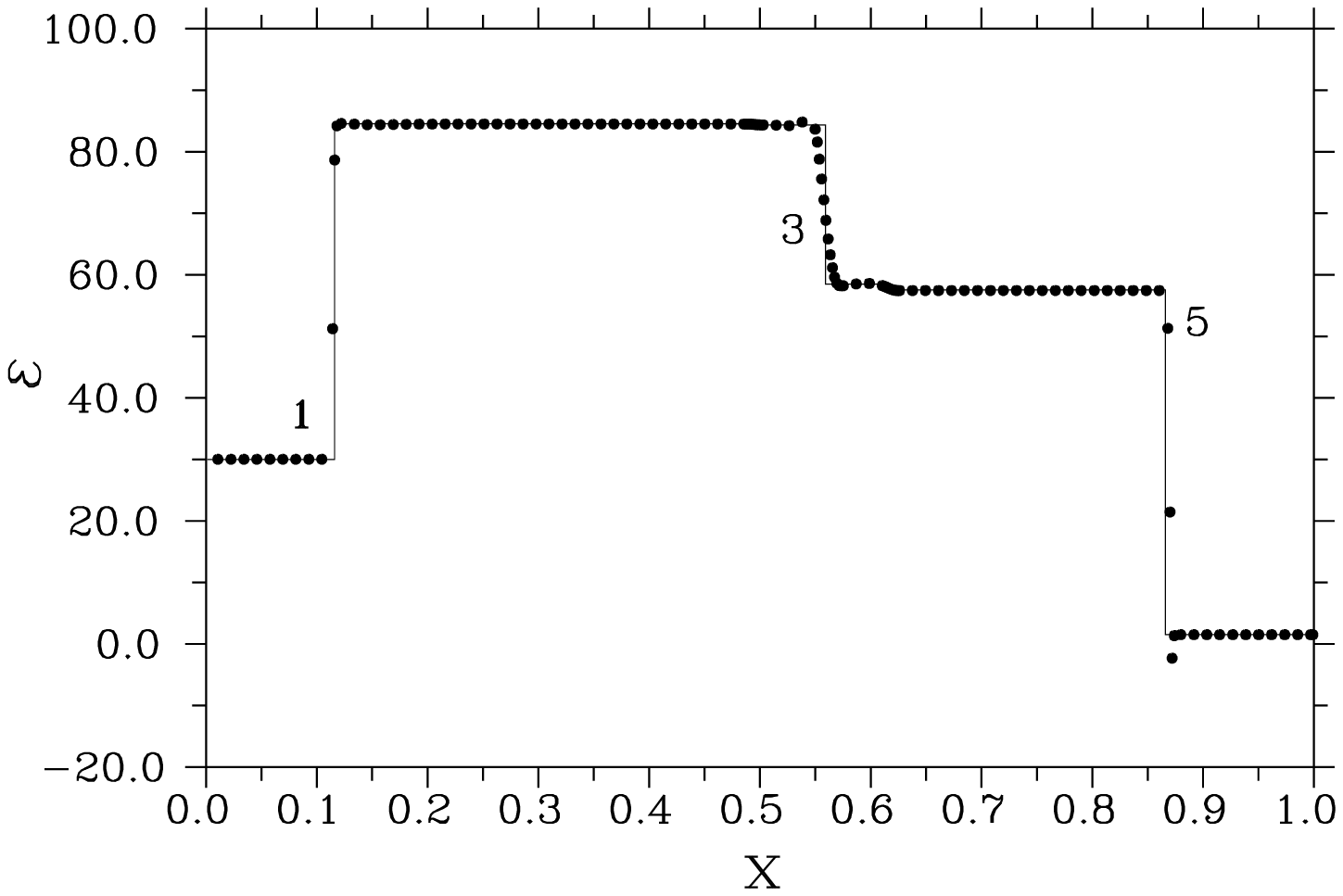,width=7cm}}}
~~\caption{Decay of an MHD discontinuity (variant 1).
The solution is corresponded to time moment $t=0.08$. The digits
correspond to: 1,5 --- fast MHD shock waves, 2 --- rotational
(alfvenic) discontinuity, 3 --- contact discontinuity, 4 --- slow
MHD shock wave}
\end{figure}

\subsubsection{Tests 3, 4. Decay of arbitrary MHD discontinuity}

The numerical simulation of an arbitrary MHD discontinuity decay
is the unique possibility to check the MHD properties of the
difference scheme. The problem statement in this test is similar
to the problem statement in test 2. The initial parameters for
this test and its exact analytical solution were taken from
paper of Ryu and Jones \cite{RyuJones1995a} (variant 1a):
$\gamma = 5/3$, $\rho_{A} = 1$, $\rho_{B} = 1$, $P_{A} = 20$,
$P_{B} = 1$, $v_{x, A} = 10$, $v_{x, B} = -10$, $B_{y,A} = 5$,
$B_{y,B} = 5$, $B_{x} = 5$. The values of another variables are
taken to be zero. This situation can arise, for example, at the
collision of two gaseous masses moving forward each to other at
the angle $\pi/4$ to the direction of the magnetic field.

The results of the numerical computation and the analytical
solution of this problem at the time $t = 0.08$ are shown on the
Fig.~4. As a result of the discontinuity decay two fast
MHD shock waves (1 and 5) arise and propagate in the opposite
directions. The rotational (alfvenic) discontinuity 2, the
contact discontinuity 3 and the slow MHD shock wave 4 are formed
between this waves. The slow MHD shock wave propagates
immediately after the fast MHD shock wave to the region of the
gas with the larger pressure.

\begin{figure}[t]
\centerline{\hbox{\psfig{figure=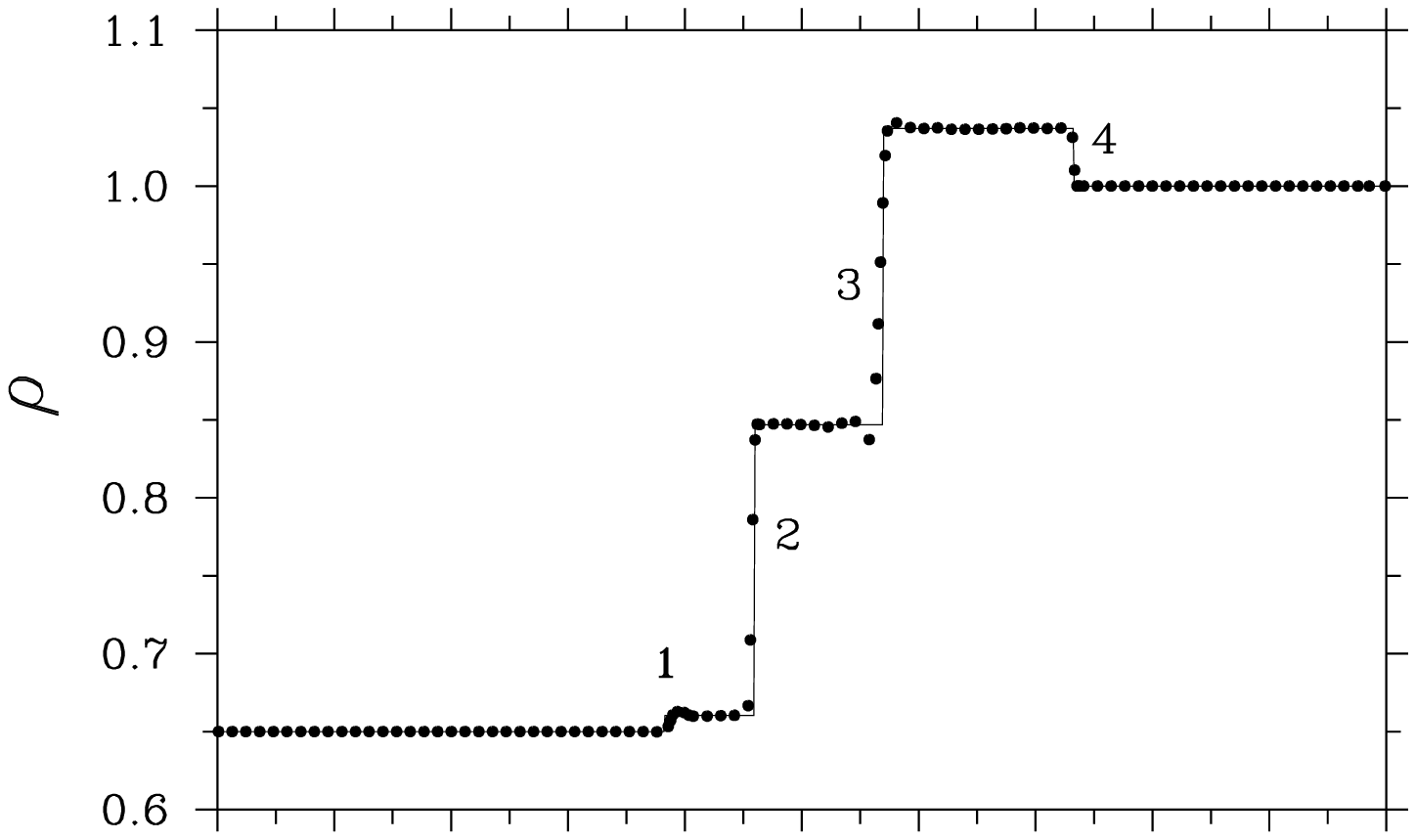,width=7cm}}
~~\hbox{\psfig{figure=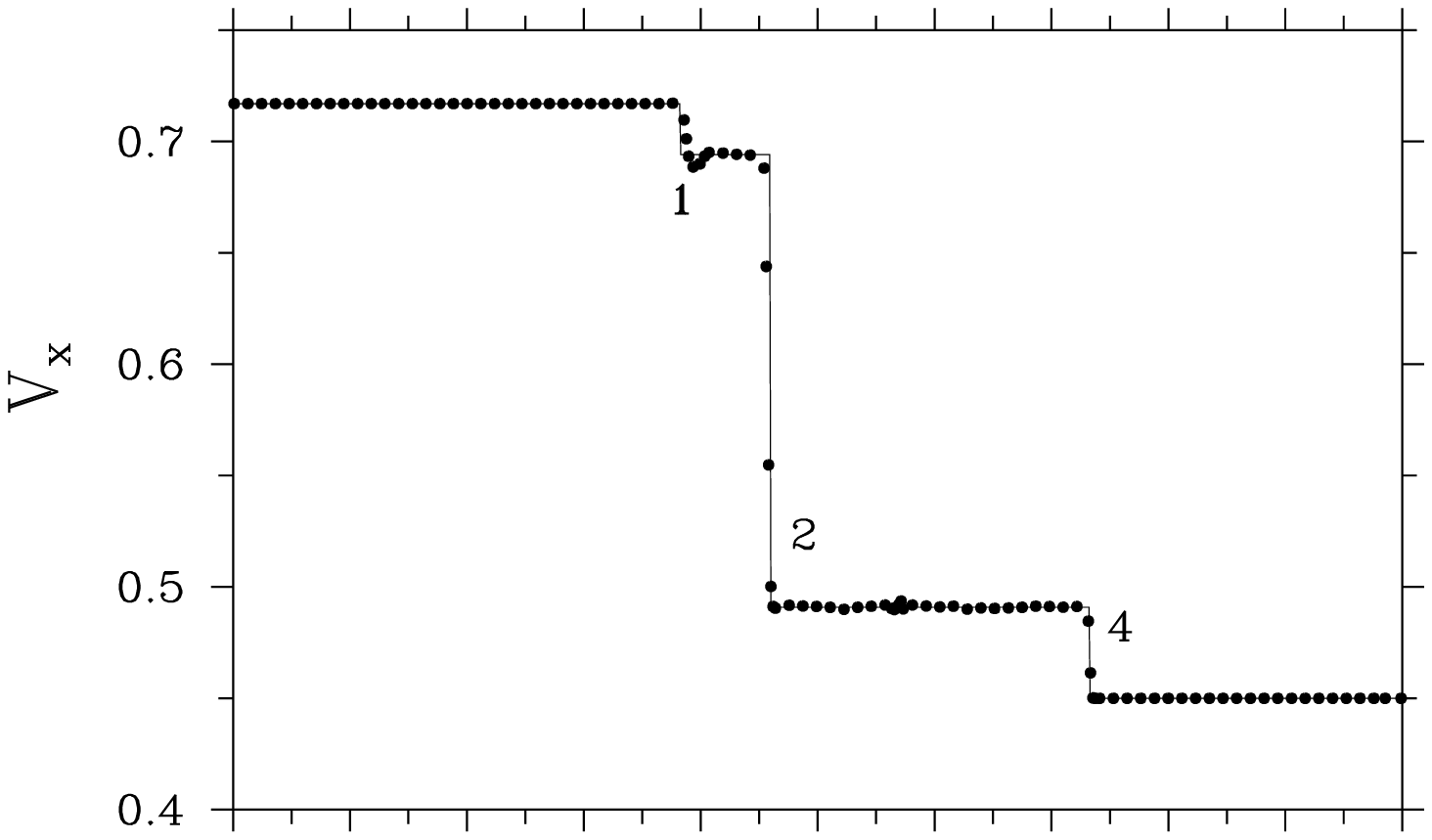,width=7cm}}}
\vspace{4mm}
\centerline{\hbox{\psfig{figure=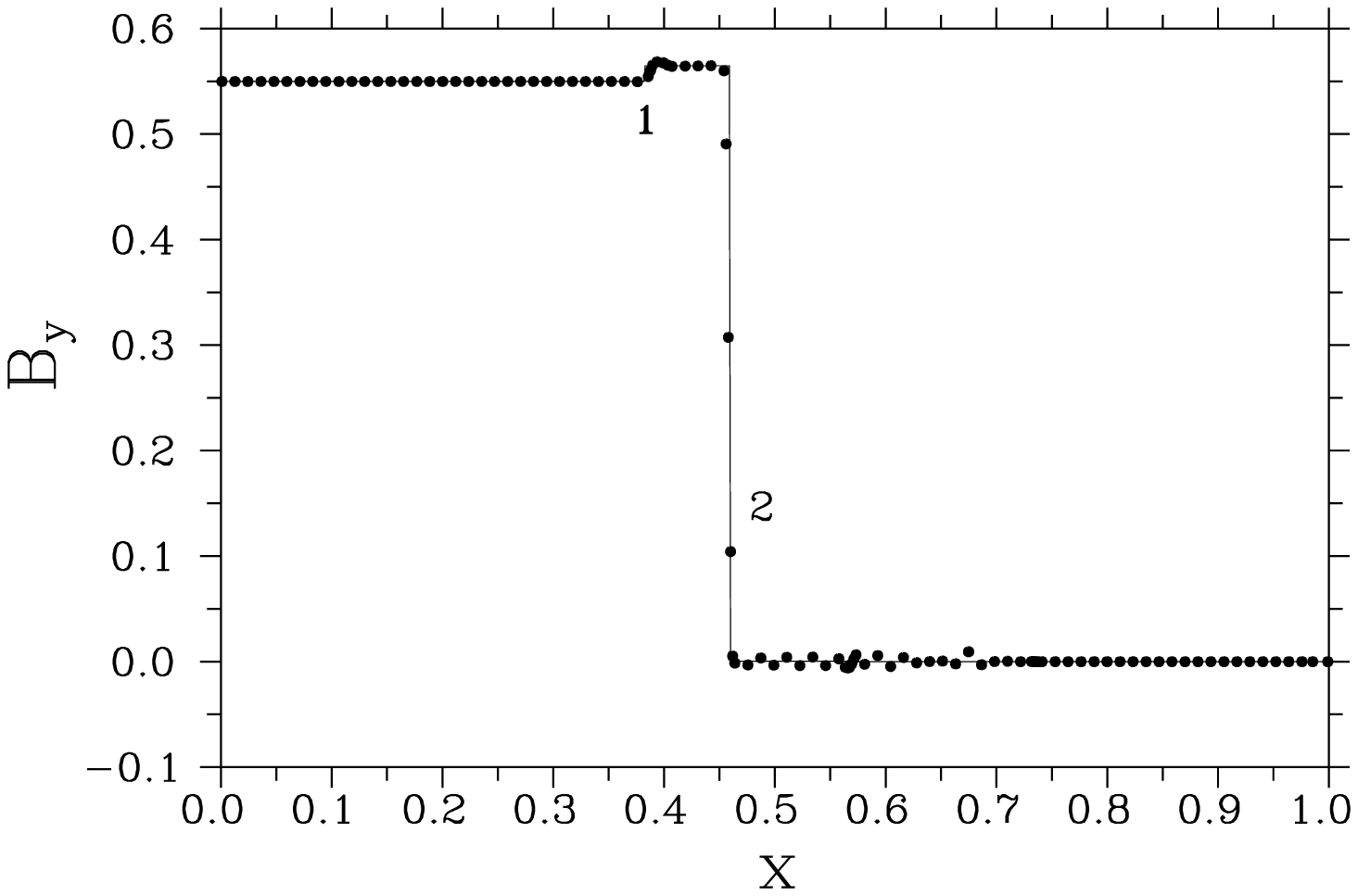,width=7cm}}
~~\hbox{\psfig{figure=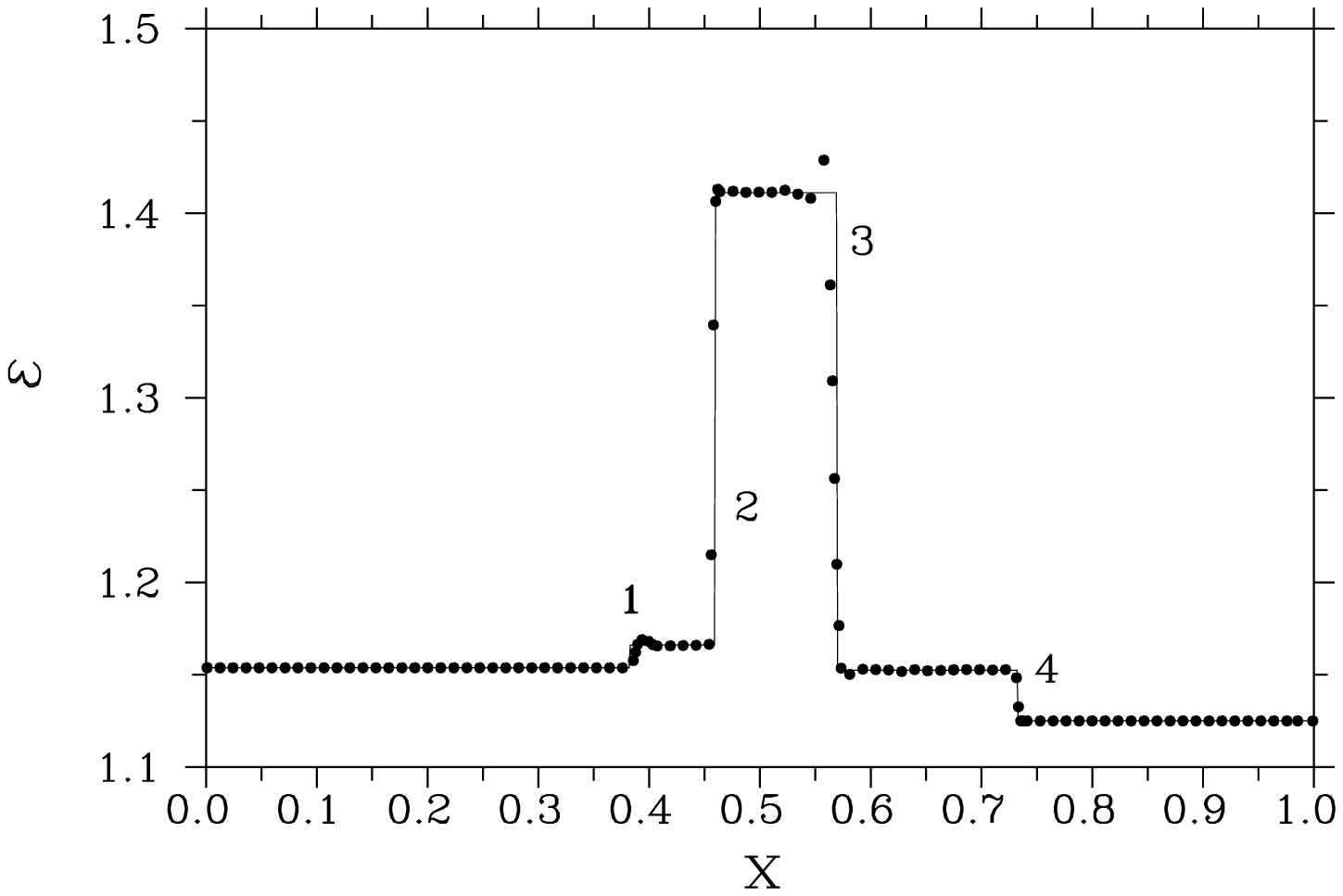,width=7cm}}}
\caption{Decay of an MHD discontinuity (variant 2).
The solution is corresponded to time moment $t=0.15$. The digits
correspond to: 1 --- fast MHD shock wave, 2 --- slow MHD shock
wave, 3 --- contact discontinuity, 4 --- gasdynamical shock
wave}
\end{figure}

In another variant of this test we consider the Riemann
problem with the initial state vector $(\rho, v_{x}, v_{y},
v_{z}, B_{y}, B_{z}, P)$ (see \cite{RyuJones1995a}, variant 4c):
region $A$ --- $(0.65, 0.667, -0.257, 0, 0.55, 0, 0.5)$, region
$B$ --- $(1, 0.4, -0.94, 0, 0, 0, 0.75)$ with $B_{x} = 0.75$ and
$\gamma = 5/3$. The switch-of slow shock wave is formed in this
case. The results of computations at the time $t = 0.15$ are shown
on Fig.~5. A fast MHD shock wave 1, switch-of slow MHD shock wave 2,
contact discontinuity 3 and gasdynamical shock wave 4 are formed.

These tests show that the scheme approximation of fast and slow
shock waves as well as the rotational (alfvenic) discontinuity
is good enough. On the contact discontinuities and on the
switch-off shock wave small oscillations occur.

\subsubsection{Test 5. Alfv\'en wave}

The problem of propagation of the finite amplitude Alfv\'en wave
(the simple alfvenic wave) is interesting because it has the
analytical solution (see \cite{LandauLifshitz}). The comparison
of the numerical and the analytical velocity profiles and the
magnetic field allows us to make conclusions about behavior of
the phase and the amplitude errors in the MHD case. We solve
the following problem. At the initial time moment the wave is
concentrated in the region $x_{L} \le x \le x_{R}$, where

\[
B_{y} = 10 \sin \left(
\pi \frac{x_{R} - x}{x_{R} - x_{L}}
-
\frac{\pi}{2}
\right)\,, \ \
B_{z} = 10 \cos \left(
\pi \frac{x_{R} - x}{x_{R} - x_{L}}
-
\frac{\pi}{2}
\right)\,,
\]
\[
v_{y} = \frac{B_{y}}{\sqrt{4 \pi \rho}}, \qquad \qquad
v_{z} = \frac{B_{z}}{\sqrt{4 \pi \rho}}\,.
\]
The wave propagates in the gas with parameters $\rho = 1$, $P =
1$, $v_{x} = 0$, $B_{x} = 1$, $\gamma = 1.4$ with velocity $a =
B_{x}/\sqrt{4 \pi \rho}$ without profile distortion. In wave
region the tangential component of the magnetic field rotates
with the conservation of its absolute value. Figure~6. shows the
numerical and analytical distributions $B_{y}$ and $B_{z}$ at
the time moment $t = 0.3$.

The comparison of these curves with each other allows us to make
the conclusion that even in the case of full MHD system the
essential smearing of the velocity and the magnetic field
profiles is absent during a sufficiently long time of the
computation. Hence the phase errors balance the amplitude ones
with the high order of accuracy.

\begin{figure}
\centerline{\hbox{\psfig{figure=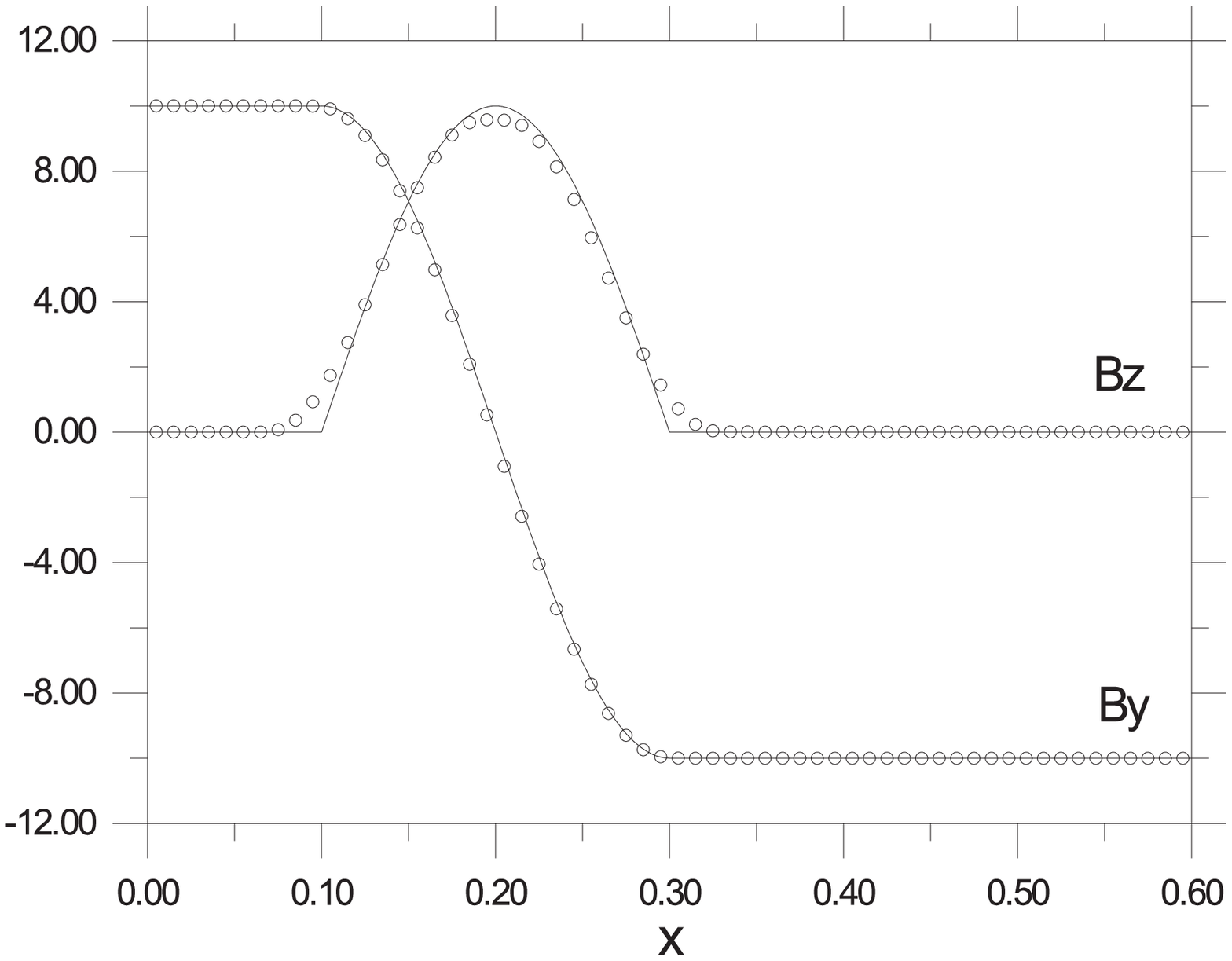,width=13cm}}}
\caption{Alfv\'en wave. The numerical (circles) and the
analytical (solid lines) distributions of the transverse
components of the magnetic field are shown at the time moment
$t=0.3$}
\end{figure}

\subsection{Two-dimensional tests}

\subsubsection{Test 6. Two-dimensional advection}

To check the diffusion properties of the developed 2D code we
obtain the numerical solution of the 2D advection
equation (see similar computations of Munz \cite{Munz1988})

\[
\frac{\partial \rho}{\partial t}
+
v_{x}(y) \frac{\partial \rho}{\partial x}
+
v_{y}(x) \frac{\partial \rho}{\partial y}
=
0
\]
in box domain $(0 \le x,y \le 1)$. The analytical solution of
this equation for the velocity profile

\[
v_{x} = -(y-y_{0})\omega\,, \qquad \qquad
v_{y} = (x-x_{0})\omega\,.
\]
describes the rotation of the density profile around the point
($x_{0}, y_{0}$) with the angular velocity $\omega$. We use in
this computation the following values: $x_{0} = y_{0} = 0. 5$,
$\omega = 1$. Hence to the time $t = 2\pi$ the original density
profile should carry out one full rotation and return to its
initial position. The computation were done on the uniform grid
with the cells number $N = 100 \times 100$. We compare the
initial state with the profile that obtained after one full
rotation.

Figure~7a shows the original density profile. It is split on two
regions. In the first region the density is equal to zero and in
the second region the density is equal to $1$. The boundary of
the regions is circle with the cut-out rectangle. The center of
the circle is not coincides with the center of the domain.
Figure~7b shows the density profile at the time $t = 2\pi$ after
one full revolution.  The profile (as in the 1D case) is smeared
onto 3--4 cells.  Therefore we can conclude that 2D variant of
our scheme has the good selfconsistent amplitude-phase
properties as well.

\begin{figure}
\centerline{\hbox{\psfig{figure=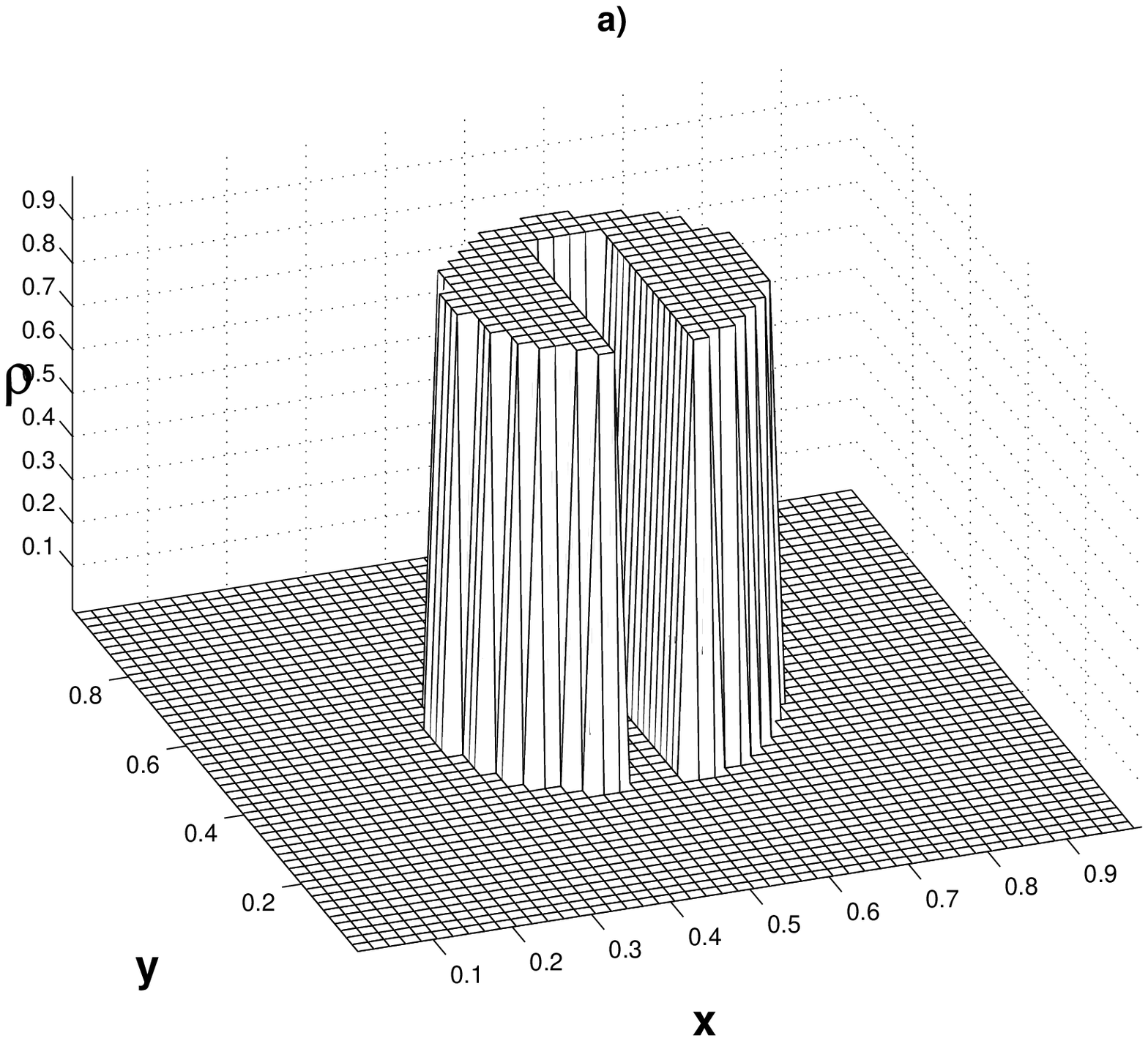,width=7cm}}
~~\hbox{\psfig{figure=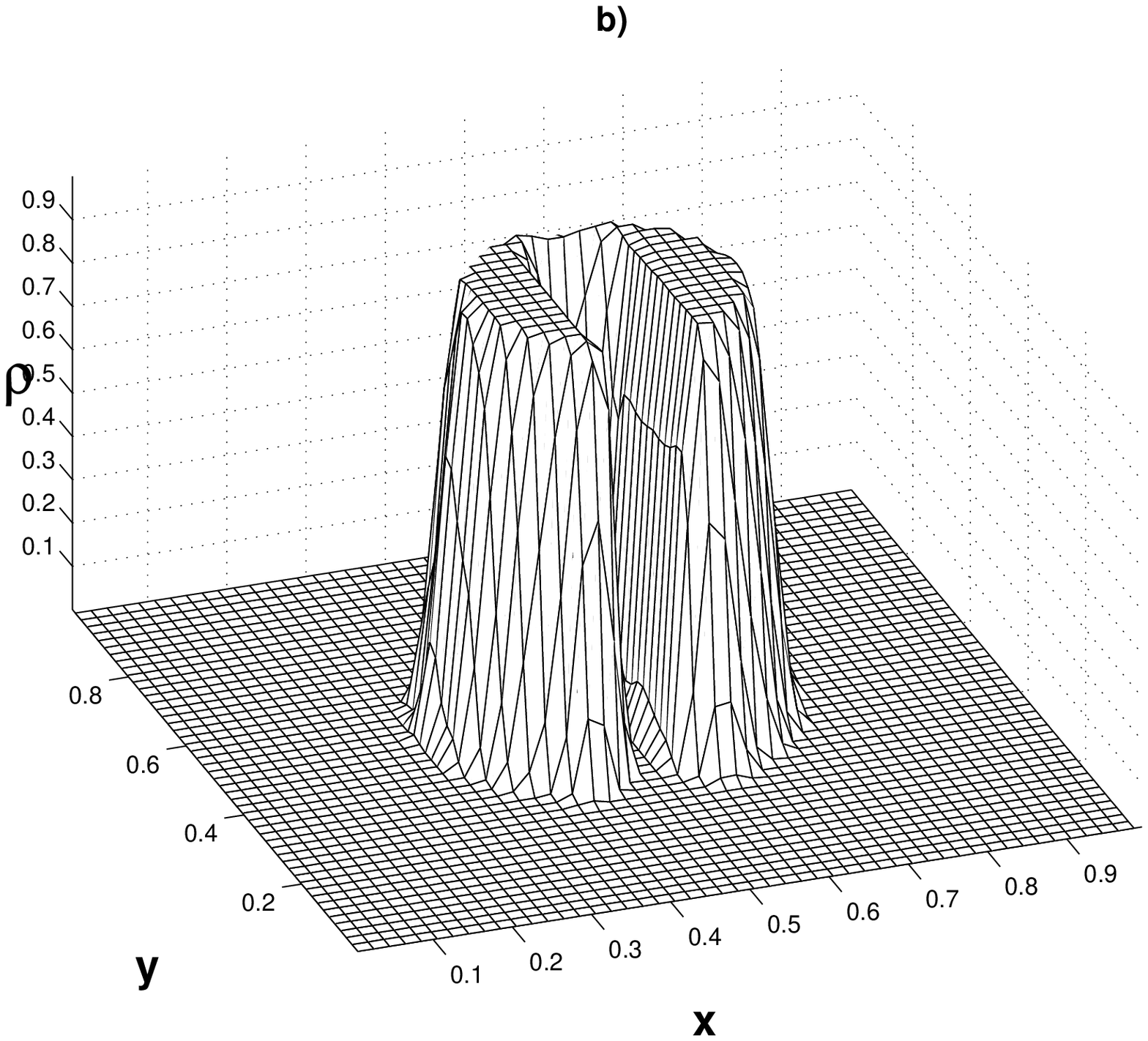,width=7cm}}}
\caption{Two-dimensional advection. The density
distributions are shown at the initial time moment (a) and after
one full revolution $t=2\pi$ (b)}
\end{figure}

\subsubsection{Test 7. Expansion of blast wave}

The test problem of blast wave expansion in the uniform medium
is very crucial for 2D numerical code due to
spherical symmetry of the wave. Simulation the blast wave
expansion allows us to check how precisely our 2D axisymmetrical
numerical code can resolve spherically-symmetrical problems.
We compute, in particular, the expansion of the supernova
remnant on the adiabatic stage. This stage is described by
Sedov--Taylor solution \cite{Sedov1946,Taylor1950,Sedov1959},
the approximation being true while the radiation cooling is
small.

We use the following initial parameters. The supernovae
explosion energy is equal to $4 \cdot 10^{50} ~\mathrm{erg}$,
the density of the interstellar medium $\rho_{0} \approx 2 \cdot
10^{-24} ~\mathrm{g} \cdot \mathrm{cm}^{-3}$, the spatial scale
of the computational domain $R_{0} \approx 4.03 \cdot 10^{19}
~\mathrm{cm}$, the pressure in the central region differs from
the pressure in the external medium at the initial time moment
approximately by eight order of magnitude. The scales of the
computational values are equal to: for distance --- $13
~\mathrm{pc}$, for velocity --- $8.9 \times 10^{8} ~\mathrm{cm}
\cdot \mathrm{s}^{-1}$, for pressure --- $1.1 \times 10^{-6}
~\mathrm{erg} \cdot \mathrm{cm}^{-3}$, for time ---
$1510~\mathrm{Yrs}$. The expansion of the supernova remnant can
be followed numerically over the time $t = 10600~\mathrm{Yrs}$.
The radius of the shock wave increases from $3.6~\mathrm{pc}$
to $12~\mathrm{pc}$ during this time.

Figure~8 shows the dependencies of the radius $R$, velocity $U$
and the pressure $P$ of the shock wave on the time. All curves
correspond to power law: $(R, U, P) \propto t^{k}$. We obtained
the following exponents: for radius of the shock wave $k_{R} =
0.403$, for velocity $k_{U} = -0.589$, for pressure $k_{P} =
-1.139$. The analytical values are $k_{R} = 0.4$, $k_{U} = -0.6$
and $k_{P} = -1.2$. We see that the numerical values of the
exponents are very close to exact Sedov--Taylor's values. The
condition of the spherical symmetry over the computations
satisfies within the accuracy of 0.3\%.

\begin{figure}
\centerline{\hbox{\psfig{figure=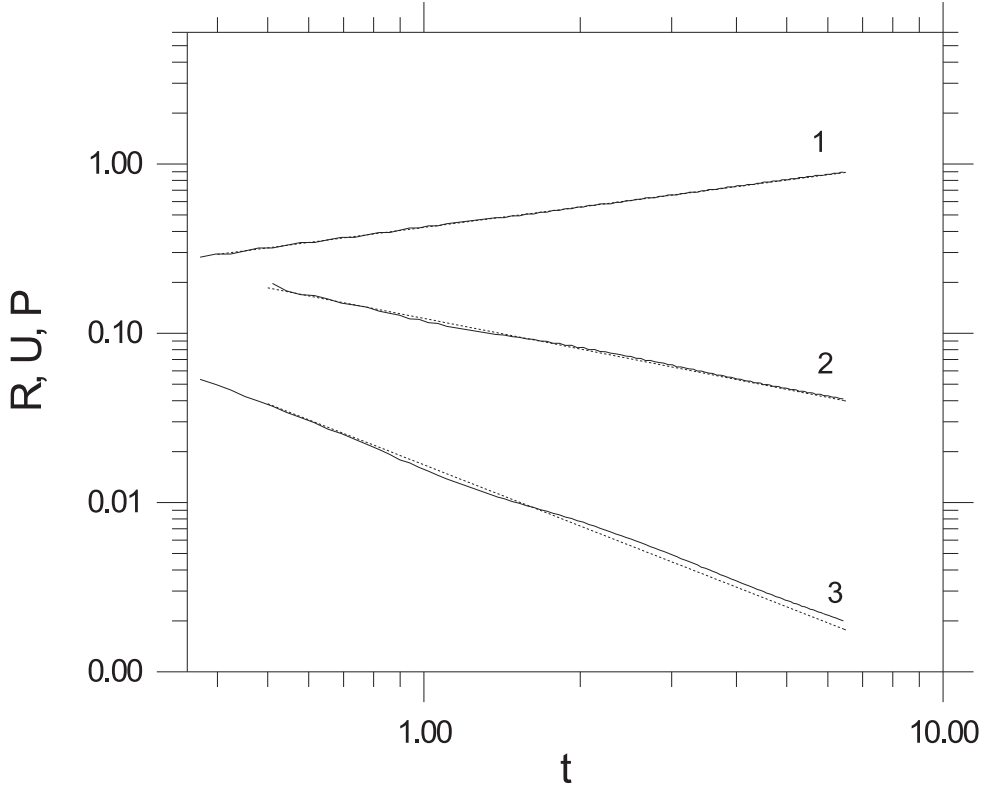,width=13cm}}}
\caption{Expansion of blast wave. The solid lines correspond
to the numerical dependencies of the radius $R$ (curve 1),
velocity $U$ (curve 2) and pressure $P$ (curve 3) of shock wave
behind its front of the time. The corresponding analytical
dependencies are shown by dotted lines}
\end{figure}

\subsubsection{Test 8. Spherically-symmetrical free-fall collapse}

For simulation of free-fall (pressure-free) collapse as an
initial state we take the uniform spherically-symmetrical cloud
with the mass $M = 1 M_{\odot}$ and with the initial density
$\rho_{0} = 10^{-19} ~\mathrm{g} \cdot \mathrm{cm}^{-3}$.
Figure~9 shows the evolution of the numerical density (dots) of
the cloud over the time in comparison with the analytical
solution (solid line). We can see that the numerical points of
density coincide with the high accuracy with values of the
analytical solution. The cloud contracts to the point at the
free-fall time $t_{ff} = \sqrt{3\pi/(32 G\rho_{0})}$. The
computation is carried out to time moment when the cloud size
decreases to the size of one grid cell. The difference of
numerical and analytical values of density is not greater than
1\% at this time and the law of the spherical symmetry
conservation satisfies within the accuracy of 0.4\%.

\begin{figure}
\centerline{\hbox{\psfig{figure=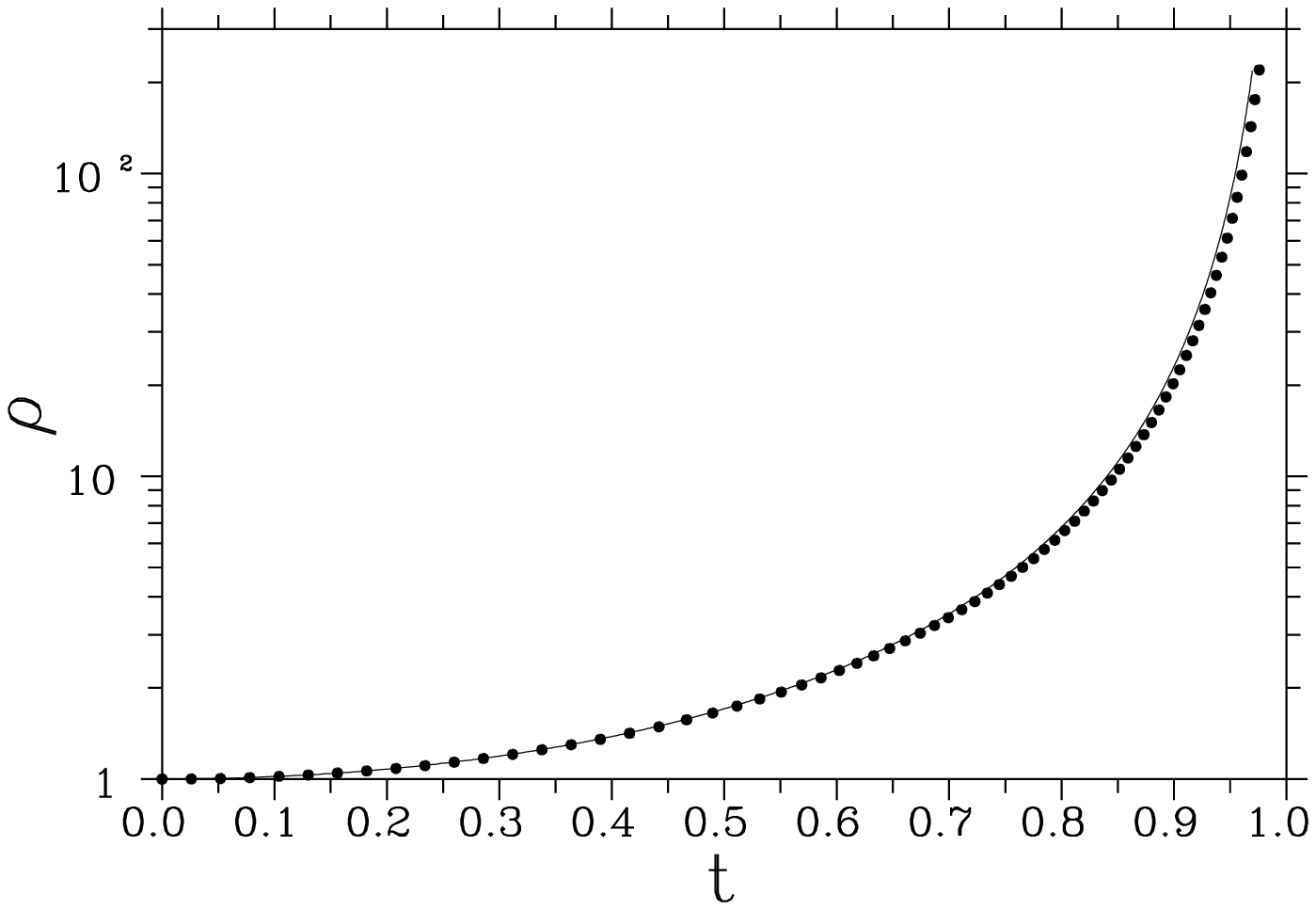,width=13cm}}}
\caption{The free-fall collapse. The numerical (dots) and
the analytical (solid line) dependencies of the cloud density of
the time are shown. As a unit of density it is used its initial
value. As a unit of time it is used the cloud free-fall time
$t_{ff}$}
\end{figure}

\subsubsection{Test 9. The contraction of polytropic cloud with
$\gamma = 4/3$}

The simulation of the selfgravitating polytropic cloud dynamics
with the adiabatic index $\gamma = 4/3$ allows us to check the
quality and the accuracy of the code in the simulation of the
selfgravitating MHD flows. The hydrostatic equilibrium of
polytropic cloud with $\gamma = 4/3$ is indifferent one since
the total energy, its first and second variations are equal to
zero (see \cite{ZeldovichNovikov1971}). The density distribution
of this cloud can be presented by analytical form as $\rho(r) =
\rho_{c} \chi^3(r/R)$, where $\rho_{c}$ is the central density,
$R$ is the cloud radius, $\chi(x) = \theta_{3}(x/\xi_{1})$
($\theta_{3}(\xi)$ is the Emden function of index 3 and $\xi_{1}
\approx 6.9$ is its first positive zero).

To check the accuracy of the conservation of mass, momentum, and
energy by our scheme we solve numerically the equations of the
gravitational gasdynamics with the initial conditions
corresponding to the specified hydrostatic state of the cloud.
We computed several models with different numbers of cells $N$
of the spatial grid. Figure~10 shows the density profiles in the
case of $N = 100 \times 100$. From the comparison of numerical
(dots) and analytical (solid line) density profiles on this
Figure we conclude that the relative error of the numerical
solution compared with the analytical one at the time $t = 1.5
t_{ff}$ is less than 2\%.  The law of total mass conservation
over the computations satisfies within the accuracy of 0.003\%
and the law of the total energy conservation satisfies within
the accuracy of 2\%.

\begin{figure}
\centerline{\hbox{\psfig{figure=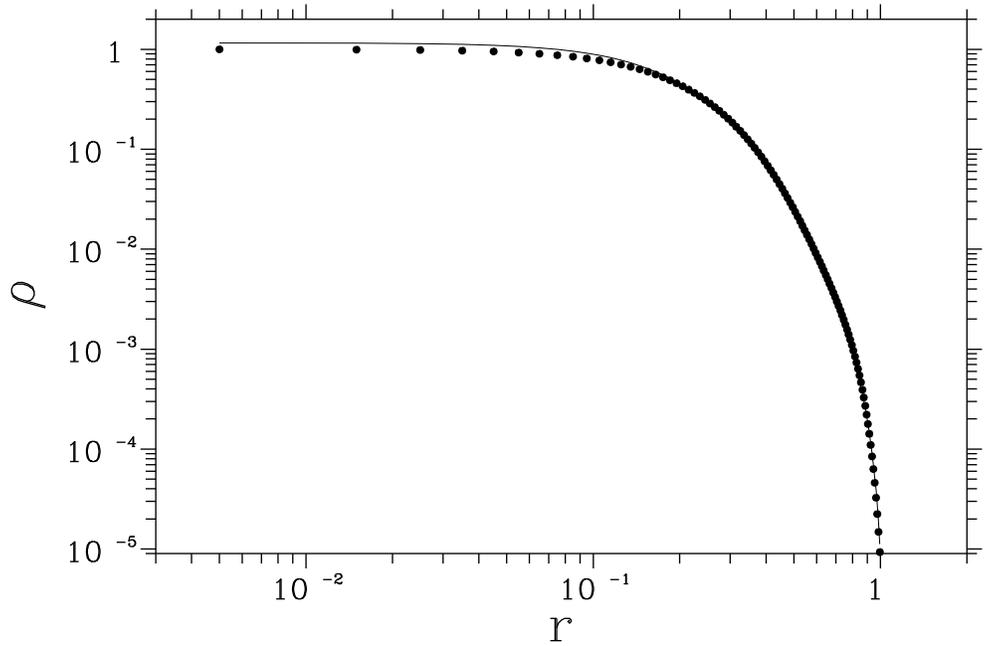,width=13cm}}}
\caption{Hydrostatic evolution of the selfgravitating
polytropic cloud with $\gamma=4/3$. The analytical (solid line)
and the numerical (dots) profiles of the density are shown at
the time moment $t = 1.5 t_{ff}$}
\end{figure}

This problem was solved on the grids with the number of cells $N
= 20 \times 20$, $80 \times 80$ and $120 \times 120$ as well. We
can say that the relative error of the numerical solution
obtained by the time $t = 1.5t_{ff}$ decreases with the
increasing of the cells number. In the case of $N = 20 \times
20$ it equals to 20\% while in the case of $N = 80 \times 80$
--- 7\%, and for $N = 120 \times 120$ it is less than one
percent. We can conclude therefore that the convergence of the
numerical solution to the analytical one with the rise of the
grid resolution takes place.

In another variant of this test we specified the deviation from
the hydrostatic equilibrium at the initial time moment by the
velocity distribution $v(r, t) = - Ur/R$. The cloud radius in
this case evolves in time in accordance with the law $R(t) =
R_{0} - U\cdot t$ and the evolution of the density and the
velocity in the cloud can be described by the following
analytical dependencies:

\[
\rho (r,t)
=
\rho_{c}(0)
\biggl (
\frac{R_{0}}{R(t)}
\chi \Bigl ( \frac{r}{R(t)} \Bigr ) \biggr )^{3} \,, \qquad \qquad
u(r,t) = - U \frac{r}{R(t)}\,.
\]
To the time $t_{0} = R_{0}/U$ the cloud will contract
to the point.

Figure~11 shows the numerical distributions of the density
(circles) and the velocity (triangles) for different times. The
solid lines correspond to analytical curves. From analysis of
this Figure we can conclude that the numerical solution with the
high accuracy coincides with the analytical one while the grid
resolution is sufficient for the adequate spatial approximation
of the central region. For the variant with $N = 100 \times 100$
this corresponds approximately to the time moment $t = 0.8
t_{0}$ (see Fig.~11).

\begin{figure}
\centerline{\hbox{\psfig{figure=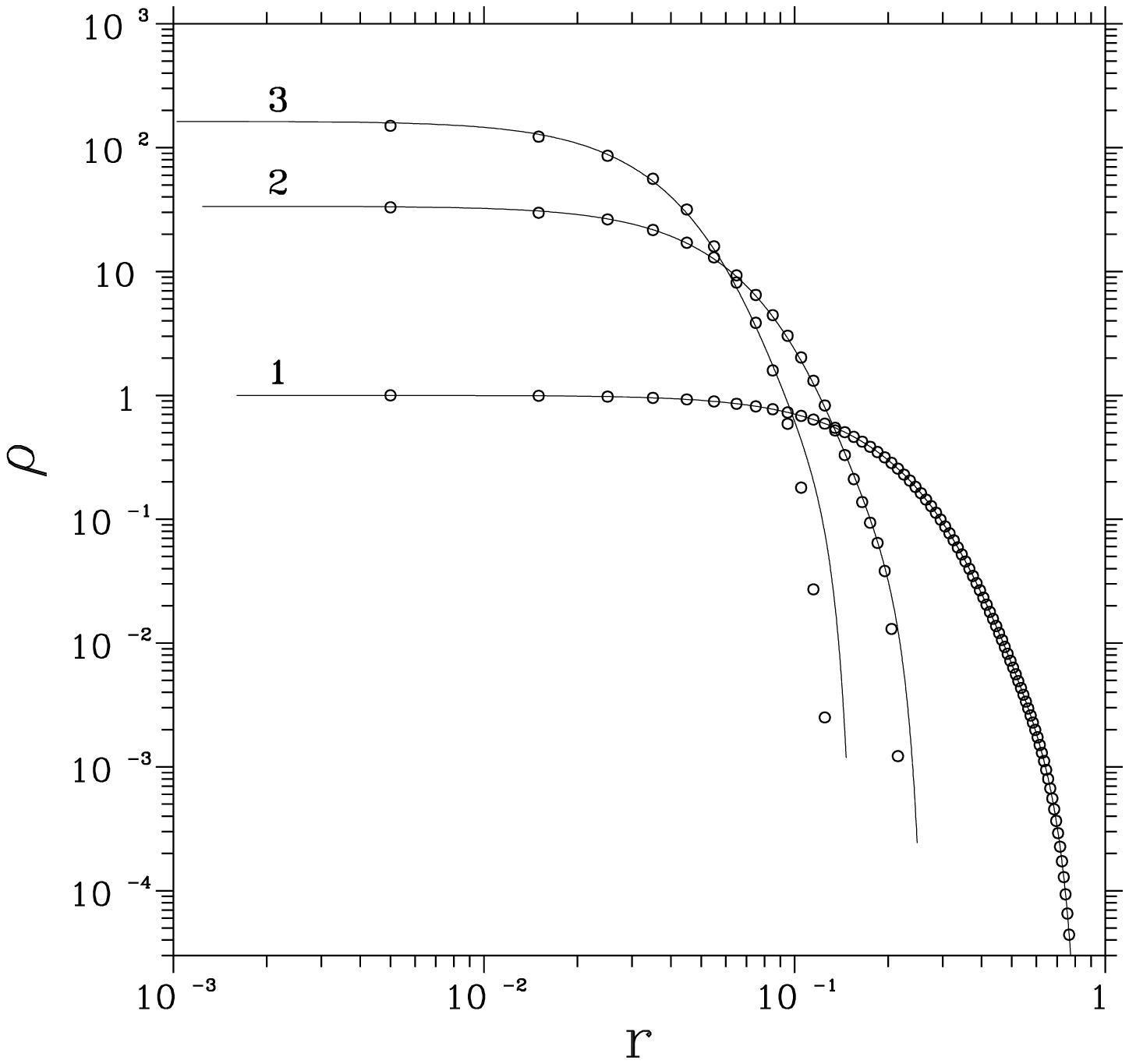,width=7cm}}
~~\hbox{\psfig{figure=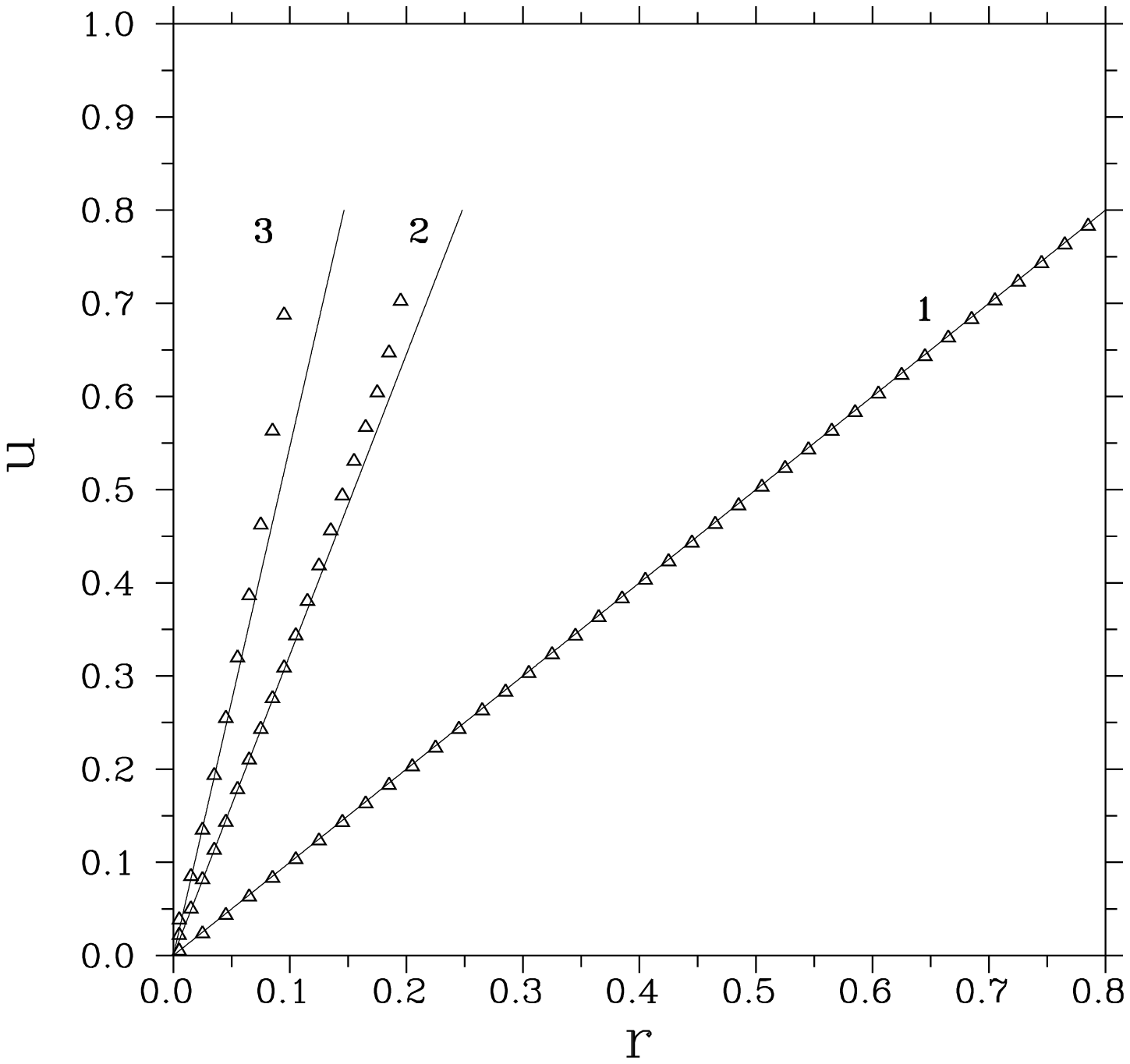,width=7cm}}}
\caption{Contraction of the polytropic selfgravitating cloud
with $\gamma = 4/3$. The distributions of the density (a) and
the velocity (b) are shown for the different moments of time.
The circles and triangles correspond to the numerical solution
and the solid lines correspond to the analytical one. The digits
1,2,3 correspond to the moments of time 0.0$t_0$, 0.69$t_0$ and
0.82$t_0$, respectively.}
\end{figure}

\section{Collapse of magnetized protostellar clouds}

\subsection{Introduction}

Protostellar clouds as the cores of interstellar molecular
clouds consist of neutral (atoms, molecules and cosmic dust
particles) and charged (electrons, ions and charged dust)
components. The equations of similar multicomponent mixture are
very complex for the numerical simulations.

The more rough ideal MHD approximation can be used as a first
approach. Contracting cloud is transparent to infrared
self-radiation on the early stages of collapse. Therefore the
initial phases of protostellar collapse can be considered as the
isothermal ones with the good approximation. This approximation
works while the optical depth of the collapsing cloud core
becomes comparable with the unity.

We performed the computation of the isothermal magnetized cloud
collapse using the uniform grid with the cells number $N = 120
\times 120$ and Courant number $C = 0.3$. We use the following
values of parameters. The initial ratio of the thermal and
kinetic rotational energies to the absolute value of the
gravitational energy $\etg = 0.386$, $\ewg = 0.0$, respectively.
The mass of the protostellar cloud $M_{0} = 1.5 M_{\odot}$ and
its temperature $T_{0} = 10 ~\mathrm{K}$. One the basis of the
relations (\ref{EQ_2014}--\ref{EQ_2017}) we can conclude that
these parameters correspond to the cloud with the initial
density $\rho_{0} = 5.18 \cdot 10^{-18} ~\mathrm{g} \cdot
\mathrm{cm}^{-3}$ and radius $R_{0} = 5.17 \cdot 10^{16}
~\mathrm{cm}$. As a result of computations we calculate the
flatness degree (ratio of the cloud thikness to its radius) and
the central density up to the free-fall time.

\subsection{Kinematics}

The simplest case of the collapse corresponds to infinitesimally
weak magnetic field. If the initial value $\emg \le 0.5 \cdot
10^{-4}$ then we can neglect the influence of the magnetic field
on the collapse dynamics. The magnetic field can be considered
as a passive admixture and for the given velocity field and may
be determined from the equation of induction (\ref{EQ_2002}).

The results of computations show that in this kinematic approach
the magnetic field $B \propto \rho^{k}$ with $k = 2/3$ since of
the cloud contraction is practically spherically-symmetrical
one.  Note that in the cloud envelope this law fulfills
approximately and in the forming core it satisfies exactly
\cite{DudorovSazonov1982}. The computations confirm the early
obtained conclusion that the magnetic field in the cloud
acquires with the time the quasi-radial geometry (see Fig.~12)
\cite{DudorovSazonov1982}. This geometry of the field becomes
distinct already to the free-fall time.

\begin{figure}
\centerline{\hbox{\psfig{figure=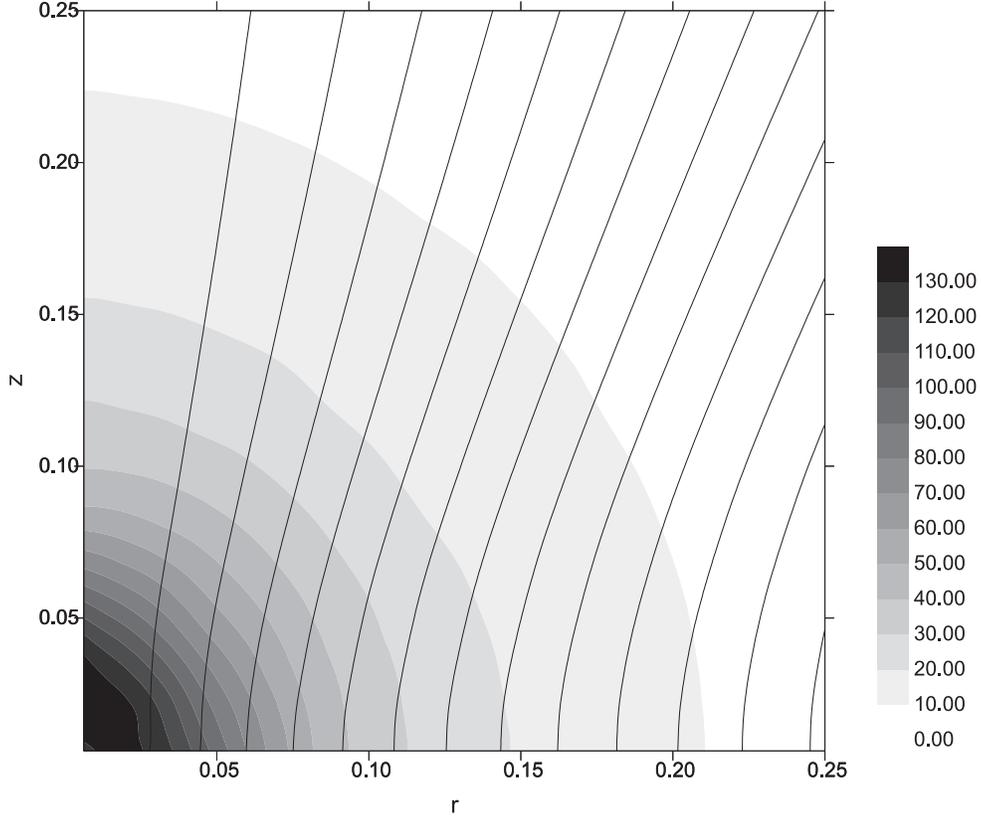,width=13cm}}}
\caption{The density distribution and the picture of the
magnetic field lines in the central region at the time moment
$t=t_{ff}$ in the case of the kinematic cloud model}
\end{figure}

\subsection{Dynamics}

In the case of intermediate magnetic fields ($0.1 \le \emg \le
0.5$) the collapse picture drastically differs from the
kinematic statement. The gas particles moving across the
magnetic field lines are deviated to the equatorial plane by the
electromagnetic forces. As a result the flat (disk-like)
structure is formed. Figure~13 shows the typical density
profiles and geometry of the magnetic field lines at the moment
of time equal to the magnetic free-fall time

\[
t_{fm} = \sqrt{\frac{3 \pi}{32 G (1 - \emg) \rho_{0}}}
\]
(see \cite{DudorovSazonov1982}). The magnetic field in the disk
is almost quasi-uniform and the ratio $B_{r}/B_{z}$ is small.

\begin{figure}
\centerline{\hbox{\psfig{figure=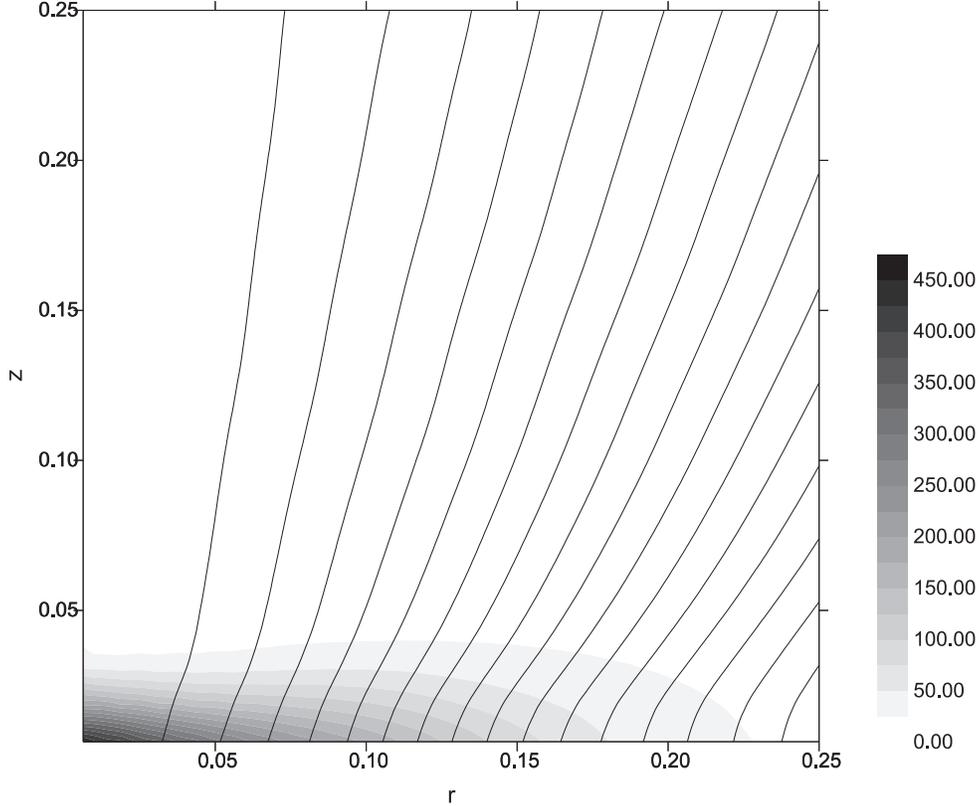,width=13cm}}}
\caption{The density distribution and lines of equal magnetic
flux in the central region of the cloud with $\emg = 0.1$ at the
time moment $t=t_{fm}$}
\end{figure}

Numerical simulations show that at the free-fall time the
flatness degree of the disk is proportional to $\emg^{-1/2}$ and
the central density is inversely proportional to $\emg$.

\subsection{Quasi-steady-state contraction}

The growth of the initial magnetic field leads to the increase
of the electromagnetic forces that support the cloud from the
contraction. Therefore at the large initial values $\emg$ the
cloud relaxes to the quasi-steady-state equilibrium. Such a
cloud should evolve within the diffusion time scale that is
determined by the ambipolar diffusion of the magnetic field. The
cloud loses the magnetic flux by the ambipolar diffusion and
slowly contracts retaining the quasi-hydrostatic equilibrium.
As a result this evolution the non-uniform profile of the
density is formed. The lifetime of such molecular cloud cores
can reach $10^{5}-10^{7}$ years.

\section{Conclusion}

The explicit conservative finite-difference TVD scheme of high
resolution for the solution of various MHD problems is
constructed. The numerical MHD code `Moon' on the basis of this
scheme is developed that can simulate the MHD flows in the 1D
and 2D approximations. The results of extensive tests show that
the code is well adapted to the simulation of many MHD problems
of plasma physics and astrophysics.

To check the dissipative properties of the developed scheme the
problems of 1D and 2D advection are solved. The results of these
tests show that the scheme smears the initial density profile
onto 3--4 cells. In the test computation of the blast wave the
condition of the spherical symmetry satisfies within the
accuracy of 0.3\%.

To analyze the convergence of the numerical solution to the
analytical one the problem of the hydrostatic equilibrium of the
selfgravitating polytropic cloud with the adiabatic index
$\gamma = 4/3$ is computed on the grids with various number of
cells. The obtained results allow us to conclude that the errors
of the numerical solution decrease to the permissible values on
the grids with the number of cells not less than $N = 100 \times
100$. In the case $N \ge 120 \times 120$ the relative error of
the numerical solution at the time moment $t = 1.55 t_{ff}$ is
less that 1\%. The numerical solution in this case coincides
with the good accuracy with the analytical one while the grid
resolution in the central region is sufficient for acceptable
approximation of the grid values.

The numerical computations of the collapse of the magnetized
protostellar cloud confirms the earlier obtained results in the
framework of 1.5D approximation (see \cite{DudorovSazonov1982}).
The magnetic field in the cloud with the initially small
intensity acquires the quasi-radial geometry after some time.
The intermediate magnetic field leads to flattening of the
collapsing clouds on the late stages of contraction. For the
strong initial magnetic field the cloud relaxes to
quasi-hydrostatic equilibrium. Such a cloud must evolve within
the diffusion time scale.

Note that we can not simulate numerically the advanced stages of
the collapse because the described code `Moon' uses the uniform
grid only. But investigation of these stages is very important
and interesting problem to understand the physics of the
formation and further evolution of young stellar objects. We are
elaborating now the adaptive mesh refinement (AMR) numerical
algorithm for the simulation of advanced stages of rotating
magnetized protostellar cloud collapse. The AMR-approach uses
the tree-threaded collection of subgrids with the subsequently
increasing grid resolution. We are constructing a new 2D MHD
numerical code `Megalion' that based on the described in this
paper high-resolution TVD scheme for the gravitational
compressible MHD flows. This code will allow us to simulate
various selfgravitating MHD flows with the large gradients and
fast time variations. New numerical code will take into account
also the processes of ambipolar diffusion, non-stationary
ionization and heating/cooling.

\begin{ack}

This work is supported partially by grants RFBR N 96--02--19005,
N 99--02--16938, N 00--02--17253, and N 00--01--10392. We wish
to thank A.V.Koldoba and G.V.Ustyugova for many informative
discussions.

\end{ack}

\end{document}